\RequirePackage{lineno}

\documentclass[preprint]{aastex}

\usepackage{amssymb}
\usepackage{amsmath}
\usepackage[applemac]{inputenc}

\newcommand{\corot}{{\textsc{CoRoT}}}
\newcommand{\cible}{{KIC7341231}}
\newcommand{\kepler}{{\textit{Kepler}}}

\newcommand{\ind}[1]{_{\rm #1}}
\newcommand{\ex}[1]{^{\rm #1}}

\def\m2s2{\,m$^{2}$\,s$^{-2}$} %m2.s -2
\def\kms{\,km\,s$^{-1}$}       %km.s -1
\def\vsini{$v\sin i$}          %vsini

\newcommand{\vaisala}{Brunt-V\"ais\"al\"a}
\newcommand{\cesam}{\textsc{cesam2k}}

\newcommand{\rchi}{\mbox{$\chi_{\nu}^2$}} 
\newcommand{\fe}{{\rm [Fe/H]}}

\newcommand{\teff}{$T\ind{eff}$}

\newcommand\T{\rule{0pt}{2.6ex}}
\newcommand\B{\rule[-1.2ex]{0pt}{0pt}}

\shorttitle{Collapsed Cores in Globular Clusters}
\shortauthors{Deheuvels et al.}

\begin{document}


\title{{\large {\sc Seismic evidence for a rapidly rotating core in a lower-giant-branch star observed with \textit{Kepler}}}}


\author{S. Deheuvels\altaffilmark{1,2,3}, R.~A. Garc{\'{\i}}a\altaffilmark{3,4}, W.~J. Chaplin\altaffilmark{3,5}, S. Basu\altaffilmark{1},  H.~M. Antia\altaffilmark{6}, T. Appourchaux\altaffilmark{7}, O. Benomar\altaffilmark{8}, G.~R. Davies\altaffilmark{4}, Y. Elsworth\altaffilmark{5}, L. Gizon\altaffilmark{9,10}, M.~J. Goupil\altaffilmark{2}, D.~R. Reese\altaffilmark{11}, C. Regulo\altaffilmark{12,13}, J. Schou\altaffilmark{14}, T. Stahn\altaffilmark{9}, L. Casagrande\altaffilmark{15}, J. Christensen-Dalsgaard\altaffilmark{16}, D. Fischer\altaffilmark{1}, S. Hekker\altaffilmark{17}, H. Kjeldsen\altaffilmark{16}, S. Mathur\altaffilmark{18}, B. Mosser\altaffilmark{2}, M. Pinsonneault\altaffilmark{3,19}, J. Valenti\altaffilmark{20}, J.~L. Christiansen\altaffilmark{21}, K. Kinemuchi\altaffilmark{22}, F. Mullally\altaffilmark{21}}

%\affil{Astronomy Department, University of California, Berkeley, CA 94720}

\altaffiltext{1}{Department of Astronomy, Yale University, P.O. Box 208101, New Haven, CT 06520-8101, USA}
\altaffiltext{2}{LESIA, UMR8109, Observatoire de Paris, Universit\'e Pierre et Marie Curie, Universit\'e Denis Diderot, CNRS, 5 Place Jules Janssen 92195 Meudon Cedex, France}
\altaffiltext{3}{Kavli Institute for Theoretical Physics, Kohn Hall, University of California, Santa Barbara, CA 93106, USA}
\altaffiltext{4}{Laboratoire AIM, CEA/DSM-CNRS-Universit\'e Paris Diderot; CEA, IRFU, SAp, centre de Saclay, 91191, Gif-sur-Yvette, France} % Garcia
\altaffiltext{5}{School of Physics and Astronomy, University of Birmingham, Edgbaston, Birmingham, B15 2TT, UK} % Chaplin
\altaffiltext{6}{Tata Institute of Fundamental Research, Homi Bhabha Road, Mumbai 400005, India} % Antia
\altaffiltext{7}{Institut d'Astrophysique Spatiale, UMR8617, Universit\'e Paris XI, B\^atiment 121, 91405 Orsay Cedex, France} % Appourchaux
\altaffiltext{8}{Sydney Institute for Astronomy (SIfA), School of Physics, University of Sydney, NSW 2006, Australia} % Benomar
\altaffiltext{9}{Institut für Astrophysik, Georg-August-Universität Göttingen, 37077 Göttingen, Germany} % Stahn, Gizon
\altaffiltext{10}{Max-Planck-Institut für Sonnensystemforschung, 37191 Katlenburg-Lindau, Germany} % Gizon, Casagrande
\altaffiltext{11}{Institut d’Astrophysique et Géophysique de l’Université de Liège, Allée du 6 Août 17, 4000 Liège, Belgium} %Reese
\altaffiltext{12}{Instituto de Astrofísica de Canarias, 38205, La Laguna, Tenerife, Spain} % Regulo
\altaffiltext{13}{Universidad de La Laguna, Dpto de Astrofísica, 38206, La Laguna, Tenerife, Spain} % Regulo
\altaffiltext{14}{W.W. Hansen Experimental Physics Laboratory, Stanford University, Stanford, CA 94305-4085, USA} % Schou
\altaffiltext{15}{Research School of Astronomy \& Astrophysics, Mount Stromlo Observatory, The Australian National University, ACT 2611, Australia} % Casagrande
\altaffiltext{16}{Danish AsteroSeismology Centre (DASC), Department of Physics and Astronomy, Aarhus University, DK-8000 Aarhus C, Denmark} % Christensen-Dalsgaard, Kjeldsen
\altaffiltext{17}{Astronomical Institute 'Anton Pannekoek', University of Amsterdam, Science Park 904, 1098 HX Amsterdam, the Netherlands} %Hekker
\altaffiltext{18}{High Altitude Observatory, NCAR, P.O. Box 3000, Boulder, CO 80307, USA} % Mathur
\altaffiltext{19}{Department of Astronomy, the Ohio State University, Columbus, OH, 43210 USA} % Pinsonneault
\altaffiltext{20}{Space Telescope Science Institute, 3700 San Martin Drive, Baltimore, MD 21218, USA} % Valenti
\altaffiltext{21}{SETI Institute/NASA Ames Research Center, Moffett Field, CA 94035}
\altaffiltext{22}{Bay Area Environmental Research Inst./NASA Ames Research Center, Moffett Field, CA 94035}

\begin{abstract}
Rotation is expected to have an important influence on the structure and the evolution of stars. However, the mechanisms of angular momentum transport in stars remain theoretically uncertain and very complex to take into account in stellar models. To achieve a better understanding of these processes, we desperately need observational constraints on the internal rotation of stars, which until very recently were restricted to the Sun. In this paper, we report the detection of mixed modes --- i.e. modes that behave both as g modes in the core and as p modes in the envelope --- in the spectrum of the early red giant \cible, which was observed during one year with the \textit{Kepler} spacecraft. By performing an analysis of the oscillation spectrum of the star, we show that its non-radial modes are clearly split by stellar rotation and we are able to determine precisely the rotational splittings of 18 modes. We then find a stellar model that reproduces very well the observed atmospheric and seismic properties of the star. We use this model to perform inversions of the internal rotation profile of the star, which enables us to show that the core of the star is rotating at least five times faster than the envelope. This will shed new light on the processes of transport of angular momentum in stars. In particular, this result can be used to place constraints on the angular momentum coupling between the core and the envelope of early red giants, which could help us discriminate between the theories that have been proposed over the last decades.
\end{abstract}


\section{Introduction}

Understanding the effects of rotation on stars is currently one of the key steps to making progress in stellar modeling. Angular momentum transport plays a central role in star and planet formation (see \citealt{bodenheimer95} and \citealt{bouvier08}, respectively). Stellar rotation is intimately linked with the star formation process. The distribution of main sequence rotation rates in stars indeed arises from the hydrodynamic assembly phase and the subsequent interaction of protostars and their accretion disks (\citealt{shu94}). Rotation also impacts stellar structure and evolution by inducing a mixing of the chemical elements inside stars. Meridional circulation is expected on general grounds in rotating stars (\citealt{eddington26}, \citealt{mestel53}, \citealt{mathis04}), and internal shears can generate hydrodynamic instabilities (see \citealt{maeder00}, \citealt{pinsonneault97} for reviews on mixing in high and low-mass stars, respectively).

Despite its importance, very little is currently known about the internal rotation of stars, the timescales over which it is modified, or the major mechanisms responsible for doing so. The theoretical problem is complex because rotation, convection, and magnetism are intricately tied to one another. Three classes of mechanisms --- hydrodynamic, wave-driven (\citealt{charbonnel05}, \citealt{mathis08}, \citealt{mathis09}), or magnetic (\citealt{gough98}, \citealt{spruit99}, \citealt{spada10}) --- could transport angular momentum or induce mixing, and their relative importance is a matter of active debate. 

To understand the mechanisms of angular momentum transport in stars better, observational data on stellar rotation are needed. Unfortunately, in most cases these are limited to observations of surface rotation. Seismology is currently the only tool that allows us to probe the internal rotation profiles of stars through the detection of rotational splittings of mode frequencies. Observations of solar p-mode splittings  have established that the convective envelope of the Sun is in mild differential rotation, whereas the radiative interior rotates as a solid body down to about 0.2 $R_\odot$ (e.g., \citealt{schou98}). This discovery revolutionized our understanding of the solar interior and raised some questions, such as the thinness of the solar tachochline (\citealt{spiegel92}), that are still open. It also showed that although rotational mixing processes have been successful in reproducing the observations for massive and intermediate-mass stars (\citealt{talon97}, \citealt{maeder00}), they predict a solar core rotation rate that is too fast (\citealt{pinsonneault89}). As a result, other more powerful transport processes, such as internal gravity waves (\citealt{charbonnel05}) or magnetic fields (\citealt{gough98}), have to be invoked.

However, the Sun alone cannot give us information about the timescale for effective angular momentum transport in stellar interiors. Indirect methods, involving the evolution of surface rotation in stellar populations, have been the only tools to study this matter so far. The spin down of sun-like stars can be used to estimate the timescale over which the radiative core responds to a magnetized wind torque.  Timescale estimates range from $\sim10$ Myr (\citealt{keppens95}) to $\sim500$ Myr (\citealt{irwin07}), with a dependence on mass and rotation rate. The survival of rapid rotation in old horizontal branch stars (\citealt{peterson83}, \citealt{behr03}) requires that they preserve a core reservoir of angular momentum, even when taking into account mass loss on the red giant branch. Differential rotation in red giants is required to explain this data (\citealt{sills00}). This was corroborated by evolutionary models of red-giant stars taking into account self-consistently the latest prescriptions of angular momentum transport (\citealt{palacios03}, \citealt{palacios06}).

In this paper, we are able to estimate the internal rotation profile of a halo early red giant star using seismology. One of the main asteroseismic goals of the space missions \corot\ (\citealt{baglin06b}) and \textit{Kepler} (\citealt{borucki10}) is to probe the rotation profile of stars. These missions are currently providing us with exquisite asteroseismic data (see \citealt{michel12} and \citealt{gilliland10} for overviews of the asteroseismology programs of \corot\ and \kepler, respectively). They have already made it possible to extract averaged seismic parameters of hundreds of solar-like stars (e.g. \citealt{2009A&A...506...41G}, \citealt{2011Sci...332..213C}, \citealt{verner11}), thousands of  red giants (e.g. \citealt{2009A&A...506..465H}, \citealt{2011A&A...525L...9M}), and also of stars in open clusters (e.g. \citealt{2010ApJ...713L.182S}). With these averaged seismic parameters, the validity of the scaling relations proposed by \cite{kjeldsen95a} could be observationally established (\citealt{huber11}) and global stellar parameters such as the mass and radius were estimated for stars in a wide range of evolutionary stages covering the HR diagram (e.g. \citealt{2009ApJ...700.1589S}, \citealt{2010A&A...522A...1K}, \citealt{2011ApJ...729L..10B}). Moreover, the excellent quality of the photometry provided by these instruments enabled us to measure individual p-mode parameters for many modes (e.g. \citealt{2010ApJ...713L.169C}, \citealt{2011A&A...530A..97B}), leading to a very precise modeling of the structure of the observed stars (e.g. \citealt{2011ApJ...740L...2S}). Observations of individual modes lead directly to observations of rotational splittings once the frequency resolution allows us to do so. 

Evolved stars such as subgiants and red giants are among the most promising objects for the purpose of studying core rotation. Indeed, these stars have higher-amplitude modes than main-sequence stars because of their higher luminosity, and more importantly, many of their non-radial modes are so-called \textit{mixed modes}. As a star evolves past the end of the main sequence, the frequencies of g modes become comparable to those of p modes owing to the high core density. When a p mode and a g mode of same degree meet, they are known to avoid each other and exchange natures, instead of simply crossing. This phenomenon, known as \textit{avoided crossing}, is caused by the coupling between the p-mode and the g-mode cavities. During this process, both modes have a mixed character: they behave  as p modes in the envelope and as g modes in the core. These mixed modes are very useful because, unlike pure g modes, they have large enough surface amplitudes so we can detect them and yet are sensitive to the structure of the core. Mixed modes have been theoretically known since \cite{osaki75} discovered them in stellar evolution models. They were first observed from the ground (\citealt{kjeldsen95b}) and later from space with \corot\ (\citealt{analyse_49385}) and \textit{Kepler} (e.g. \citealt{campante11}, \citealt{2011ApJ...733...95M}). These modes enabled us to probe the structure of the core of subgiants (\citealt{deheuvels11}, \citealt{metcalfe10}) and red giants (\citealt{2011Sci...332..205B}, \citealt{2011A&A...532A..86M}), thus making it possible to discriminate between evolutionary scenarios (\citealt{2011Natur.471..608B}, \citealt{mosser12}, see \citealt{bedding12} for a review). The detection of mixed modes that are split by rotation will allow us  to probe the rotation rate of stars even in their deepest interior (\citealt{kawaler99}, \citealt{lochard04}). Very recently, \cite{beck12} were able to measure the rotational splittings of mixed modes in three red giant stars observed with \textit{Kepler} and they concluded that the core must rotate at least ten times faster than the surface in these objects. 

We here report on the detection of rotationally-split mixed modes in the oscillation spectrum of the star \cible, lying on the lower giant branch and observed with the \textit{Kepler} spacecraft. We use these modes to probe the internal rotation profile of this star. The rest of the paper is organized as follows: we first give an overview of the atmospheric parameters of the star in Sect. \ref{sect_atm}. We analyze in detail the oscillation spectrum of the star, obtained from one year of \textit{Kepler} data, in Sect. \ref{sect_sismo}. This spectrum is made very complex by the presence of many mixed modes. We show that the observed non-radial modes are rotationally split and that the rotational splitting varies from one mode to the other, which we interpret as a possible evidence for radial differential rotation in the star. To study this hypothesis, we search for a stellar model that is in good agreement with both the atmospheric and the seismic constraints of the star in Sect. \ref{sect_model}. Finally, in Sect. \ref{sect_inversion}, we use the observed rotational splittings and our closest stellar models to infer information about the rotation of the star. In particular, we perform inversions of the rotational profile using both the Regularized Least Squares method and the Optimally Localized Averages technique.

\section{Atmospheric parameters \label{sect_atm}}

\subsection{Existing measurements}

\begin{table*}
  \begin{center}
  \caption{Atmospheric parameters of \cible\ found in the literature or derived in this study. \label{tab_spectro}}
\begin{tabular}{c c c}
\hline \hline
\T \teff\ (K) & $5470\pm150$ & Present study (IRFM, Sect. \ref{sect_photo}) \\
 & $5233\pm50\, [100]^\sharp$ & Present study (HIRES spectroscopy, Sect. \ref{sect_spectro}) \\
 & $5915\pm182$ & \cite{molenda08} \\
 & $5483$ & \cite{ammons06} \\
 & $6000$ & \cite{latham02} \\
 & $5300$ & \cite{cayrel01} \\
 & $5251$ & \cite{thorburn94} \\
\B & $5301$ & \cite{laird88} \\
\hline
\T \fe\ (dex) & $-1.4\pm0.1$ & Present study (Strömgren calibration, Sect. \ref{sect_photo}) \\
 & $-1.64\pm0.05\,[0.1]^\sharp$ & Present study (HIRES spectroscopy, Sect. \ref{sect_spectro}) \\
 & $-1.68\pm0.19$ & \cite{molenda08} \\
 & $-0.79$ & \cite{ammons06} \\
 & $-1.51$ & \cite{pilachowski93} \\
\B & $-2.18$ & \cite{laird88} \\
\hline
\T $\log g$ & $3.55\pm0.03$ & Present study (seismology, Sect. \ref{sect_obs}) \\
 & $4.06\pm0.29$ & \cite{molenda08} \\
 & 3.7 & \cite{charbonnel05} \\ 
\B & 4.0 & \cite{latham02} \\ 
\hline
\T \B \vsini\ (\kms) & $<1.0\pm0.5\, [1]^\sharp$ & Present study (HIRES spectroscopy, Sect. \ref{sect_spectro}) \\
\hline
\end{tabular}
\\ {\small $^\sharp$ The errors in brackets are obtained by multiplying the internal errors by a factor two (see text). These values were used in the present study.}
\end{center}
\end{table*}

The star \cible\ is also known as HIP92775 or G205-42. It has a $V$ magnitude of 9.96 (\citealt{laird88}). The extremely high proper motion ($39.18\pm0.85$ mas/yr in RA and $255.25\pm1.24$ mas/yr in DE, \citealt{vanleeuwen07}) and radial velocity ($-269.16\pm0.14$ \kms, \citealt{latham02}) make it an unambiguous halo star. 

The existing measurements of the star's atmospheric parameters are listed in Table \ref{tab_spectro}. The star was found to be very metal-poor relative to the Sun --- values quoted in the literature range from $-2.18$ dex (\citealt{laird88}) to $-0.79$ dex (\citealt{ammons06}). Because of this, it has been included in several catalog studies of low-metallicity stars. However, the atmospheric parameters found by these works are often not consistent between each other. For instance, there is a wide spread in the determination of the effective temperature. Most values range from $5470$ K (\citealt{casagrande10}) to 5483 K (\citealt{ammons06}), but we also note that two studies found a significantly higher effective temperature, around 6000 K (\citealt{latham02}, \citealt{molenda08}). These latter measurements are however dubious because they are both found along with a $\log g$ around 4.0 for the star, which is completely inconsistent with the seismic value of the surface gravity that we derive in the present work ($\log g=3.55\pm0.03$, see Sect. \ref{sect_obs}).
 
%Strömgren photometry for the star : (b-y)=0.425, m1=0.098, c1=0.223 (schuster06) + infrared photometry from 2mass catalogue (skrutskie06) => IRFM method using the polynomials given by casagrande10 gives a temperature of $T\ind{eff}=5470\pm150$ K. good agreement with the photometric estimate of ammons06
%metallicity : [Fe/H]$=-1.4\pm0.1$ dex (casagrande11)

%$\log g$ Silva-Aguirre 2011: 3.53 (IRFM table), Charbonnel 2005 : 3.7, (Molenda Zakowicz 2008 : 4.06) \\

%$T\ind{eff}$ IRFM table (Casagrande 2010) : 5091, Thoburn 1994 : 5251, Laird, Carney, Latham 1988 : 5301, (Molenda Zakowicz 2008 : 5915) \\

%[Fe/H] Pilakowski 1993 : -1.51, Laird, Carney, Latham 1988 : -2.18, (Molenda Zakowicz 2008 : -1.68) \\

%The star is evolved, probably in the post main sequence stage. Very under-metallic.

\subsection{Photometric determination of the star's $T\ind{eff}$ and [Fe/H] \label{sect_photo}}

Strömgren photometry is available for this star. By using the values of $(b-y)$, $m_1$, and $c_1$ that were obtained by \cite{schuster06} for this star along with the Strömgren metallicity scale proposed by \cite{casagrande11}, we derived an estimate of the metallicity of $[$Fe/H$]=-1.4\pm0.1$ dex, confirming that the star is metal-poor. We also combined the Tycho2 $B_T V_T$ photometry (\citealt{hog00}) with the infrared $JHK\ind{S}$ photometry available from the 2MASS catalogue (\citealt{skrutskie06}) and applied the InfraRed Flux Method (IRFM) method as prescribed by \cite{casagrande10} to derive an estimate of the effective temperature of the star. We thus obtained $T\ind{eff}=5470\pm150$ K. To determine the error bar on this measurement, we took into account the uncertainty on the star's metallicity and surface gravity, although the IRFM is only weakly dependent on these quantities. The largest sources of uncertainty come from the reddening (depending on the reddening calibration that we use, we obtain $0<E(b-y)<0.02$) and the errors in the photometric observations.	
\begin{figure}
\begin{center}
\includegraphics[width=8cm]{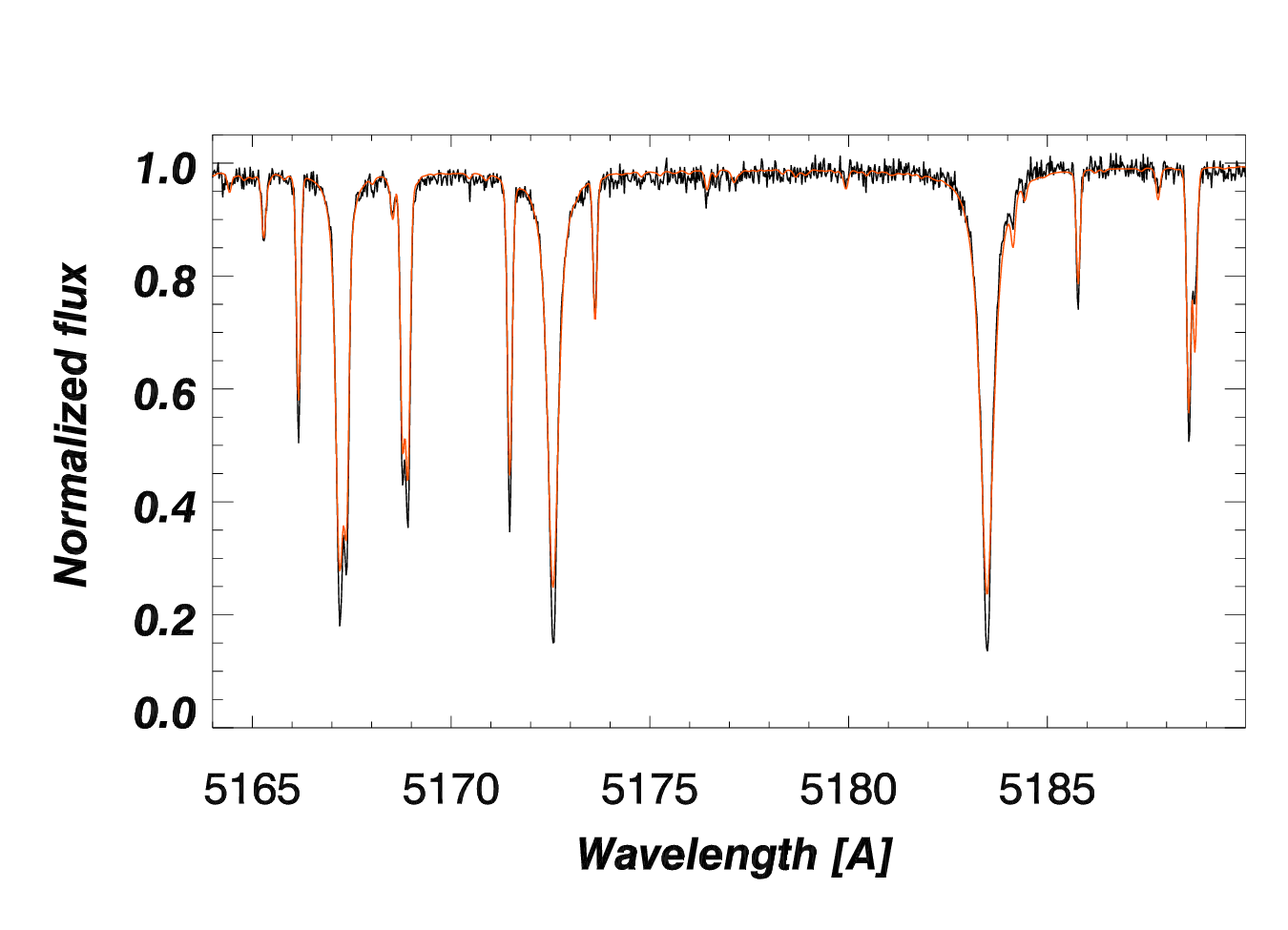}
\end{center}
\caption{Spectrum of \cible\ obtained with the HIRES spectrograph at the Keck Observatory around the Mg B triplet lines. The red dashed line corresponds to the best-fit model.
\label{fig_spectro}}
\end{figure}

\subsection{A new spectroscopic determination of the atmospheric parameters of the star \label{sect_spectro}}

Because of the wide variations between the atmospheric parameters found for \cible\ by previous studies, we re-observed the star using the HIRES spectrograph (\citealt{vogt94}) at the Keck Observatory on 26 July 2011 in order to perform a dedicated spectroscopic analysis. The spectrum we obtained has a SNR of about 100 and the B5 decker was used, yielding a spectral resolution, R = 52,000. The analysis of the spectrum was carried out using Spectroscopy Made Easy (SME), a spectral synthesis modeling program (\citealt{valenti96}, \citealt{valenti05}). The 1-D LTE radiative transfer code built into SME was used to fit the continuum-normalized spectrum of the star. 

It is well known that the determination of $\log g$ from spectroscopy alone is not reliable (e.g. \citealt{smalley05}). Instead, global seismic properties of the star (large separation of p modes and frequency of the maximum amplitude) provide a significantly more precise estimate of the surface gravity through well-established scaling laws (\citealt{kjeldsen95a}) which have recently been further validated from observations (\citealt{huber11}). For our analysis, we fixed the surface gravity to the seismic value of $\log g$ that we obtain from the \textit{Kepler} oscillation spectrum in the present work ($\log g=3.55$, see Sect. \ref{sect_obs}). This has the advantage of further constraining the other spectroscopic parameters such as the effective temperature or the metallicity.

We searched for optimal values of \teff, \vsini, and \fe, using a Levenberg-Marquardt algorithm to minimize the \rchi\ parameter between the model and observed spectrum. After obtaining one set of model parameters, we varied the effective temperature by $\pm 400$ K as input to new SME trials to explore the parameter degeneracy. In each case, the best-fit model returned to the values listed in Table \ref{tab_spectro}, within half the value of the stated uncertainties. A wavelength segment of the spectrum of the star, spanning the Mg B triplet lines, is shown in Fig. \ref{fig_spectro}. The model fit is overplotted as a red dashed line in this plot. %\textbf{The depths of some lines in our synthetic profile slightly differ from those of the observed spectrum --- the depth of several lines are underestimated over the segment plotted in Fig. \ref{fig_spectro}. The cores of deep lines are indeed formed in NLTE, while SME is an LTE code. However, the information about pressure broadening and surface gravity are contained in the line wings. The deep line cores are thus masked out when fitting the lines (see ).}

We note that this star is very metal poor and it is quite possible that systematic sources of error exceed our estimates of the formal errors. For example, the stellar atmosphere model used to generate the synthetic spectrum was obtained by interpolation from a grid of Kurucz model atmospheres, which may not be robust for such a metal poor star. 
%\textbf{The $\alpha$-elements are probably more abundant than in solar metallicity stars, which could be the reason why the depths of several lines are underestimated by our synthetic profile (see Fig. \ref{fig_spectro}).} 
Furthermore, the atomic line data (oscillator strengths and broadening coefficients) were originally tuned to match the high resolution solar spectrum. To account for these likely sources of systematic errors in our analysis the formal internal errors have been multiplied by an arbitrary factor of two to provide the uncertainties listed in Table \ref{tab_spectro}. We thus obtained an effective temperature of $T\ind{eff}=5233\pm100$ K, lower than but in marginal agreement with our result from photometry. In the following, we chose to use the photometric estimate of $T\ind{eff}$ as a reference instead of the spectroscopic measurement because the former is directly tied to the fundamental definition of the effective temperature. However, we have checked that all our results remain unchanged when considering the spectroscopic temperature instead.
%In the following, we adopted a value of the temperature of $T\ind{eff}=5350\pm180$ K, by combining our photometric and spectroscopic measurements.}

\section{Seismic properties of \cible \label{sect_sismo}}

\subsection{\textit{Kepler} observations \label{sect_obs}}

The star \cible\ was observed with the \textit{Kepler} spacecraft over a period of one year (quarters\footnote{The \kepler\ spacecraft operates over 4 seasons (quarters) each year. At the beginning of each quarter, data from the previous quarter are downlinked to the ground, new target tables are uploaded, and the spacecraft is rotated by 90 degrees to reposition the solar arrays.} Q5, Q6, Q7, and Q8 of \textit{Kepler} observations) with the short-cadence mode (58.84876 s), allowing the detection of solar-like oscillations. The lightcurve was processed using the \kepler\ mission data pipeline (\citealt{jenkins10}) and corrected for outliers, occasional jumps and drifts following \cite{garcia11}. We also corrected the lightcurve for long-period instrumental drifts by applying a high-pass filter (subtraction of a triangularly smoothed version of the lightcurve over a width of 1 day). 

\begin{figure}
\begin{center}
\includegraphics[width=8cm]{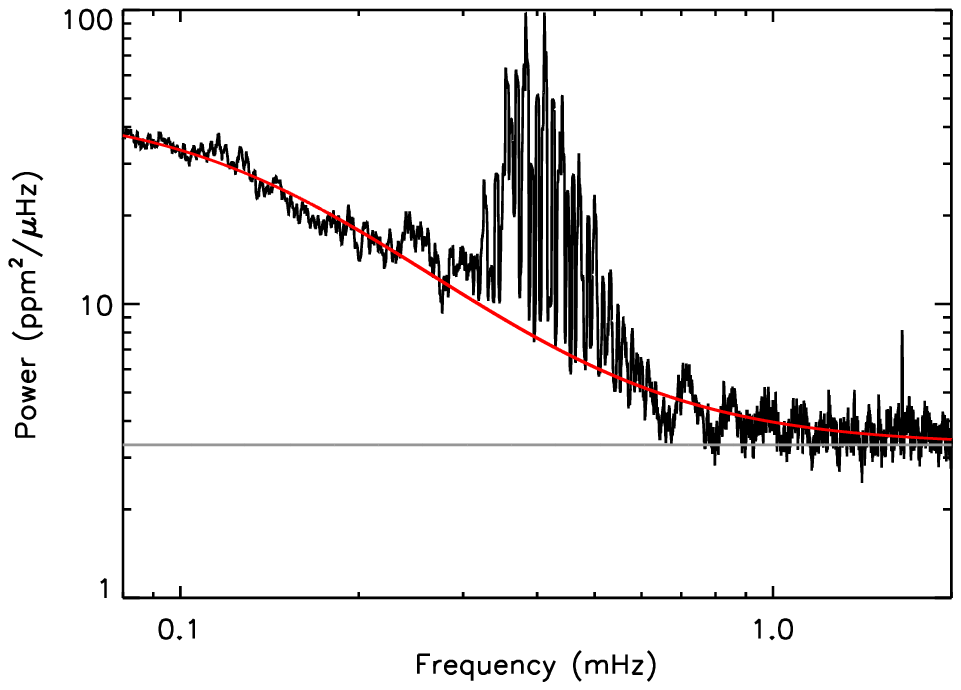}
\includegraphics[width=8cm]{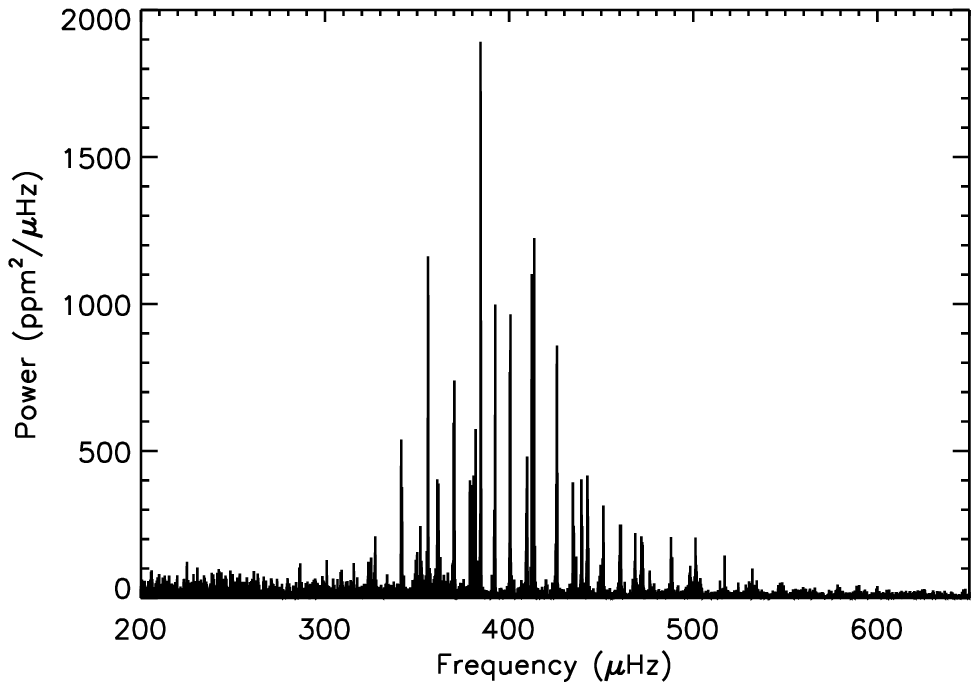}
\end{center}
\caption{Power density spectrum of \cible\ computed with one year of \textit{Kepler} observations. \textbf{Left:} Spectrum smoothed over a 2-$\mu$Hz boxcar. The red curve corresponds to a fit of a function of the type $B(\nu)$ (see text) to the spectrum. The contribution from the photon noise is overplotted in gray. \textbf{Right:} Power density spectrum over the frequency range where oscillation modes are observed.
\label{fig_spec}}
\end{figure}

The oscillation spectrum of the star (shown in Fig. \ref{fig_spec}) was obtained from the time series by using the Lomb-Scargle periodogram (\citealt{scargle82}). It exhibits a clear excess of power between 300 and 600 $\mu$Hz, which corresponds to the signature of high-order pressure modes. The frequency of the maximum signal is $\nu\ind{max}=406\pm3\,\mu$Hz (this value is estimated in Sect. \ref{sect_mle}). Using the scaling relations suggested by \cite{brown91}, we can derive an estimate of the surface gravity from $\nu\ind{max}$ and $T\ind{eff}$. Using our photometric estimate of the effective temperature ($T\ind{eff}=5470\pm150$ K, see Sect. \ref{sect_photo}), we obtained a seismic estimate of $\log g=3.55\pm0.03$. We note that even when considering instead our spectroscopic estimate of the effective temperature ($T\ind{eff}=5233\pm100K$, see Sect. \ref{sect_spectro}), our estimate of $\log g$ is not significantly modified (we then obtain $\log g=3.54$).
%the higher surface temperature obtained by \cite{molenda08}, our estimate of $\log g$ is not significantly modified (we then obtain $\log g=3.57$).

\begin{figure*}
\begin{center}
\includegraphics[width=12cm,clip=1]{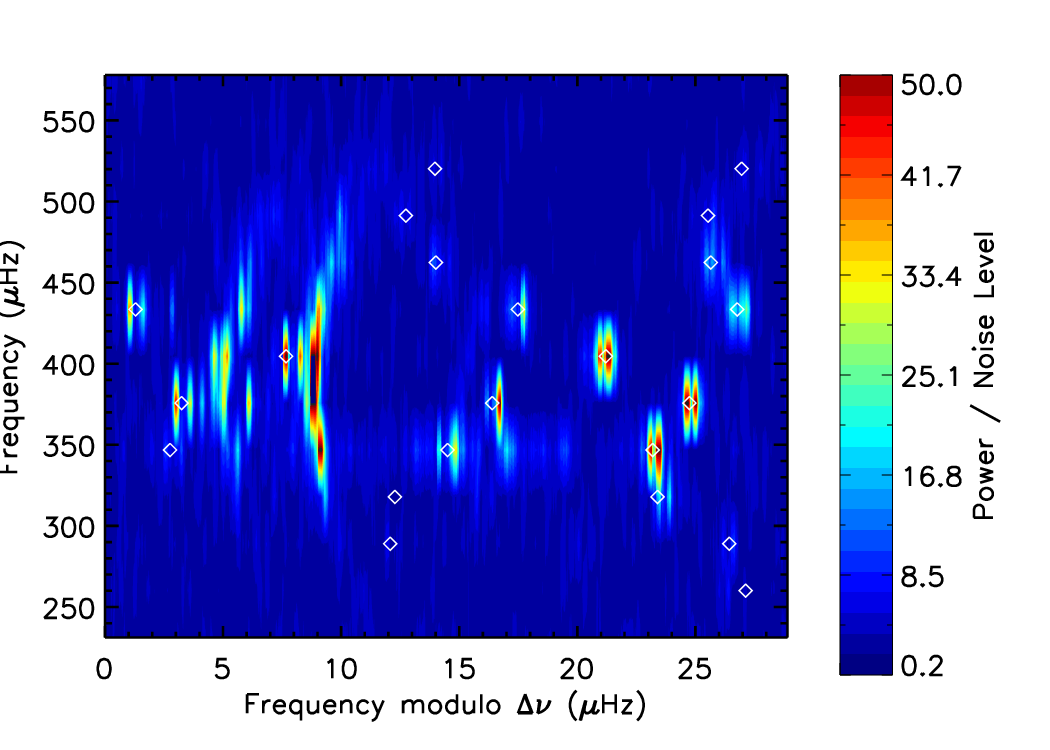}
\end{center}
\caption{\'Echelle diagram of \cible\ folded with a large separation of $\Delta\nu=29\,\mu$Hz. For more clarity, the power spectrum was binned over a 0.2-$\mu$Hz boxcar and clipped at a maximum of 50 times the noise level. The white diamonds correspond to the mode frequencies predicted by the method of \cite{benomar12} (see text).
\label{fig_echobs}}
\end{figure*}

An autocorrelation of the power spectrum provides a first estimate of the mean large separation $\langle\Delta\nu\rangle\sim29\,\mu$Hz. Fig. \ref{fig_echobs} shows an \'echelle diagram of the observed power spectrum folded with this value of the large separation. Two clear neighboring ridges stand out (at an abscissa around 7 $\mu$Hz and 3 $\mu$Hz in Fig. \ref{fig_echobs}), which are readily identified as the $l=0$ and $l=2$ ridges, respectively. No actual $l=1$ ridge can be identified in Fig. \ref{fig_echobs}. Instead of lying along a ridge, the $l=1$ modes appear scattered in the \'echelle diagram. This is caused by the fact that these modes all have a mixed behavior and therefore no longer follow the asymptotic approximation of high-order p modes. \cite{rome} have shown that $l=1$ avoided crossings induce distortions in the whole $l=1$ ridge, owing to the strong coupling between the p-mode and the g-mode cavities for these modes. The existence of numerous $l=1$ g modes in the frequency range of the observation can therefore account for such a scattered $l=1$ ridge. This hypothesis is corroborated by the low value of $\log g$ found for this star, which indicates that it is probably in the post main sequence stage. 

We note that in this case, we also expect the $l=2$ modes to have a mixed behavior. However, $l=2$ p modes are known to propagate less deeply inside the star than $l=1$ p modes. As a result, the evanescent zone that couples the p-mode and the g-mode cavities is wider for $l=2$ modes and the coupling is therefore much weaker. This causes the $l=2$ modes to be more efficiently trapped. In this case, the modes that are trapped in the core have a much larger inertia, and thus a much longer lifetime, than the modes trapped in the envelope (\citealt{dziembowski01}). For stars at the bottom of the red giant branch, \cite{dupret09} found a lifetime of the order of $10^3$ to $10^4$ days for $l=2$ modes trapped in the core. Even with one year of \textit{Kepler} data, these modes are thus expected to be unresolved. This is important because \cite{dupret09} showed that the heights of unresolved modes (i.e. their maximum power spectral densities) are inversely proportional to their inertia. As a result, the $l=2$ modes that are trapped in the core should have very small heights. This explains why we detect only the $l=2$ modes that are trapped mainly in the envelope, i.e. those that are located near the $l=2$ ridge of p modes in the \'echelle diagram.

One particularly striking feature of the oscillation spectrum is that the $l=1$ modes seem to be split in two peaks. This phenomenon can even be seen with the naked eye in Fig. \ref{fig_echobs}. This is clearly the signature of stellar rotation with a high inclination angle between the rotation axis and the line of sight. It opens very exciting possibilities because many of the non-radial modes have a mixed nature. The modes that behave mainly as g modes are more sensitive to the rotation in the core, whereas the modes trapped in the acoustic cavity are more sensitive to the rotation in the envelope. With \cible, we thus have the opportunity to probe the internal rotation profile and look for possible signatures of differential rotation in radius. We note that a detailed observation of the lightcurve did not show any clear stellar modulation produced by the crossing of spots at the surface of the star, although the inclination angle is expectedly high. This is compatible with an old star without any strong magnetic field in the outer layers that could trigger surface activity.

\subsection{Identification of the modes \label{sect_id}}

First, we need to identify the $l=1$ mixed modes in the observed power spectrum. \cite{benomar12} prescribed a method to identify the modes in oscillation spectra with avoided crossings by using a toy model proposed by \cite{rome}. If the p and g-mode cavities were uncoupled, their eigenmodes would coexist independently and no avoided crossings would occur (\citealt{aizenman77}). We will further denote as $\pi$ modes the theoretical uncoupled p modes and as $\gamma$ modes the uncoupled g modes. To interpret the observed spectrum, which can be very complex because of mixed modes, it is very convenient to consider the uncoupled modes as harmonic oscillators and to introduce a coupling term between them to simulate the effects of the evanescent zone (\citealt{rome}). Indeed, this toy model translates into a simple eigenvalue problem that can readily be solved to estimate the eigenfrequencies of the coupled system. These frequencies can then be fitted to the observed ones (central points of the observed rotational multiplets), by adjusting the frequencies of the theoretical $\pi$ modes, those of the $\gamma$ modes and the strength of the coupling (\citealt{benomar12}). The convergence of such a method is made easier by taking into account the fact that the uncoupled modes roughly follow the asymptotic approximation of \cite{tassoul80} ($\pi$ modes are approximately equally spaced in frequency and $\gamma$ modes in period). %One tricky point is the number of $\gamma$ modes that need to be introduced. The more evolved the star is, the more numerous these modes are.

In the present case, we found a satisfactory fit to the observations by considering 15 $l=1$ $\gamma$ modes between 310 and 580 $\mu$Hz, with a mean period spacing of $\langle\Delta\Pi_1\rangle=107.1\pm2.3$ s, and a set of $\pi$ modes with a mean large separation of $\langle\Delta\nu\rangle=28.9\pm0.2\,\mu$Hz. The frequencies of the $l=1$ modes that we obtained using this simple model are overplotted in Fig. \ref{fig_echobs}. We thus confirm that most of the modes detected outside of the $l=0$ and $l=2$ ridges can be identified as $l=1$ mixed modes and we obtain a list of guessed frequencies for the observed $l=1$ modes. In a complementary way, we analyzed the mixed modes using the asymtotic relation that was applied to hundreds of red-giant stars by \cite{mosser12}. We obtained estimates of the $l=1$ mixed mode frequencies that are very close to the ones found with the method of \cite{benomar12}. We derived a period spacing of $112.8\pm0.3$\,s for the $l=1$ modes, in marginal agreement with the value obtained by using the method of \cite{benomar12}.

Unfortunately, these two methods can not be applied to $l=2$ modes. Indeed, the coupling between the cavities is too weak and as mentioned above, only the modes that are close to acoustic modes are detected. However, knowing that the period spacing of $l=2$ modes is such that $\Delta\Pi_2 = \Delta\Pi_1 / \sqrt{3}$, the asymptotic relation of \cite{mosser12} allow us to predict that $l=2$ avoided crossings should occur around the $\pi$ modes of radial orders $n$=11, 12, 13 (frequencies around 327, 355, 384\,$\mu$Hz). We come back on this matter in Sect. \ref{sect_model}, using best-fit models of the star. %For instance, around a frequency of $380\,\mu$Hz... Rely on a model?

\subsection{Estimating the rotational splittings of the observed non-radial modes \label{sect_mle}}

Rotation is known to lift the degeneracy between the non-radial modes of same radial order $n$ and degree $l$ but different azimuthal order $m$. In the case where the rotation of the star is slow enough so that the effects of the centrifugal force can be neglected, a first-order perturbation approximates well the effects of rotation on the mode frequencies. The star we study here is in this case because it has a low projected surface velocity ($v\sin i <1\pm1$ \kms) and a high inclination angle (this latter point will be confirmed in Sect. \ref{sect_extract}), which places it significantly below the limit of validity of the perturbative approach (see e.g. \citealt{suarez10}). If we further assume that the rotation profile is spherically symmetric, the frequency of the $(n,l,m)$ mode is given by
\begin{linenomath*}
\begin{equation}
\nu_{n,l,m}=\nu_{n,l,0}+m\delta\nu_{n,l}
\label{eq_split_th}
\end{equation}
\end{linenomath*}
where $\delta\nu_{n,l}$ is known as the \textit{rotational splitting} and can be expressed as a weighted measure of the star's rotation rate $\Omega(r)$
\begin{linenomath*}
\begin{equation}
\delta\nu_{n,l} = (2\pi)^{-1} \int_0^R K_{n,l}(r) \Omega(r) \,\hbox{d}r
\label{eq_split_inv}
\end{equation}
\end{linenomath*}
The functions $K_{n,l}(r)$ are the \textit{rotational kernels} of the modes; they depend on the equilibrium structure of the star and on the mode eigenfunctions. The expression of rotational splittings for spherically symmetric rotating stars, which is  given by Eq. \ref{eq_split_th}, was first obtained by \cite{cowling49} and \cite{ledoux49} (for a review on the effect of rotation on the mode frequencies, see e.g. \citealt{goupil11}). We note that Eq. \ref{eq_split_th} implies that the components of a rotational multiplet are expected to be uniformly spaced by the splitting $\delta\nu_{n,l}$.

To estimate the rotational splittings of the modes in the oscillation spectrum of the star, we followed and adapted a procedure that was designed to analyze the oscillation spectra of solar-like pulsators observed from space (\citealt{fletcher06}, \citealt{appourchaux06}), which was successfully applied to \corot\ targets (e.g. \citealt{analyse_49933}, \citealt{analyse_49385}) and \textit{Kepler} targets (e.g. \citealt{campante11}).
%\subsubsection{$l=1$ modes}

\subsubsection{Modeling the Power Spectral Density (PSD)}

The observed power spectrum is distributed around a mean profile $P(\nu)$, following the statistics of a $\chi^2$ with 2 degrees of freedom (\citealt{duvall86}). The profile $P(\nu)$ can be split into two components: 
\begin{itemize}
\item the background $B(\nu)$, which is composed of the photon noise and the contribution from granulation. The modeling of $B(\nu)$ is described in Appendix \ref{app_bg},
\item the contribution from the stellar pulsations to the PSD, which is described below.
\end{itemize}

Solar-like oscillations are stochastically excited by the turbulent motions in the outer convective envelope. They are intrinsically damped and their profiles in the PSD can be modeled as a Lorentzian function with a linewidth inversely proportional to the mode lifetimes (\citealt{duvall86}). We can thus write
\begin{linenomath*}
\begin{equation}
F(\nu) = \sum_{n,l,m} \frac{a_{l,m}(i)H_{n,l}}{1+4[\nu-\nu_{n,l}+m\delta\nu_{n,l}]^2/\Gamma_{n,l}^2},
\label{eq_lorentz}
\end{equation}
\end{linenomath*}
where $\nu_{n,l}$, $H_{n,l}$ and $\Gamma_{n,l}$ correspond to the frequency, height, and linewidth of the $m=0$ component of the $(n,l)$ multiplet. Inside a rotational multiplet, the height ratios only depend on the inclination angle $i$ of the star (\citealt{gizon03}) and correspond to the terms $a_{l,m}(i)$ in Eq. \ref{eq_lorentz}. Unlike previous analyses of that type, here we allow the rotational splitting $\delta\nu_{n,l}$ to vary from one mode to the other. The peaks are indeed narrow enough so we can measure individual splittings, and hence obtain observational constraints on the variations in the rotation rate with radius.

Several assumptions that are commonly made while analyzing the spectra of main-sequence solar-like pulsators are no longer expected to be valid in the case of a star as evolved as this one. For instance, we usually assume that in each overtone, all the modes share a common linewidth $\Gamma$. Here, numerous non-radial modes have a strongly mixed behavior, which affects their inertia and therefore their lifetimes. We thus considered the linewidths of all the modes as free parameters of our fit. Similarly, for main-sequence stars, the ratio between the height of a non-radial mode and the height of the closest radial mode is usually fixed to a theoretical value, determined from the stellar limb-darkening profile (see e.g. \citealt{gizon03}). In the present case, we know that some modes are mainly trapped in the core and have longer lifetimes. We can therefore not expect all the modes to be resolved, which forbids us to use these theoretical visibility ratios. All the mode heights are thus left free in our fit.

\subsubsection{Extraction of the mode parameters \label{sect_extract}}

The mode parameters have then been independently determined by seven teams, either by using the maximum likelihood estimation (MLE) method (see \citealt{anderson90} for more details on how this method is applied to oscillation spectra) or by performing a maximum a posteriori (MAP) estimation based on Bayesian priors (e.g. \citealt{gaulme09}). Generally, a simultaneous fit of all the modes is preferred because numerous parameters can be considered to be common to several modes (e.g. linewidths, height ratios...). However, in our case, most of these assumptions had to be abandoned and the only parameter which is shared by the modes is the inclination angle of the star. For this reason, several teams chose to perform local fits of the modes to gain in flexibility, leaving the inclination angle free for each mode in the fit.

%\paragraph{$l=1$ modes}

For $l=1$ modes, the fitting teams obtained estimates of all the mode parameters (including the rotational splittings) in close agreement with one another. In order to obtain a robust list of rotational splittings, we selected among the fitted $l=1$ modes a subset of 15 modes for which at least six of the seven teams agreed on all parameters within 1-$\sigma$ error bars, as prescribed by \cite{analyse_49933}. For these modes, the fits all converged toward a high inclination angle (between 70$^{\circ}$ and 90$^{\circ}$). This was expected because the overall profile of the $l=1$ modes (which appear to be split in two prominent peaks) can only be explained by $i\sim90^{\circ}$.

%\paragraph{$l=2$ modes}

The case of $l=2$ modes is more complicated. For a rotation axis perpendicular to the line of sight, the $m=\pm1$ components of the $l=2$ multiplets vanish and we thus expect to detect rotationally-split $l=2$ modes as triplets composed of the $m=0$ and $m=\pm2$ components. In practice, the profiles of $l=2$ modes are clearly split by rotation only in the three radial overtones around the maximum of the signal. Outside of this region, the signal-to-noise ratio is too low to detect the signature of rotation. For these latter modes, there was a poor agreement between the different teams concerning the rotational splittings and we thus chose to discard them in the following. In two of the three overtones around $\nu\ind{max}$, the $l=2$ modes (around 409 and 439 $\mu$Hz) exhibit the expected behavior and their rotational splittings were successfully estimated (see Table \ref{tab_splittings}). 

\begin{figure}
\begin{center}
\includegraphics[width=8cm]{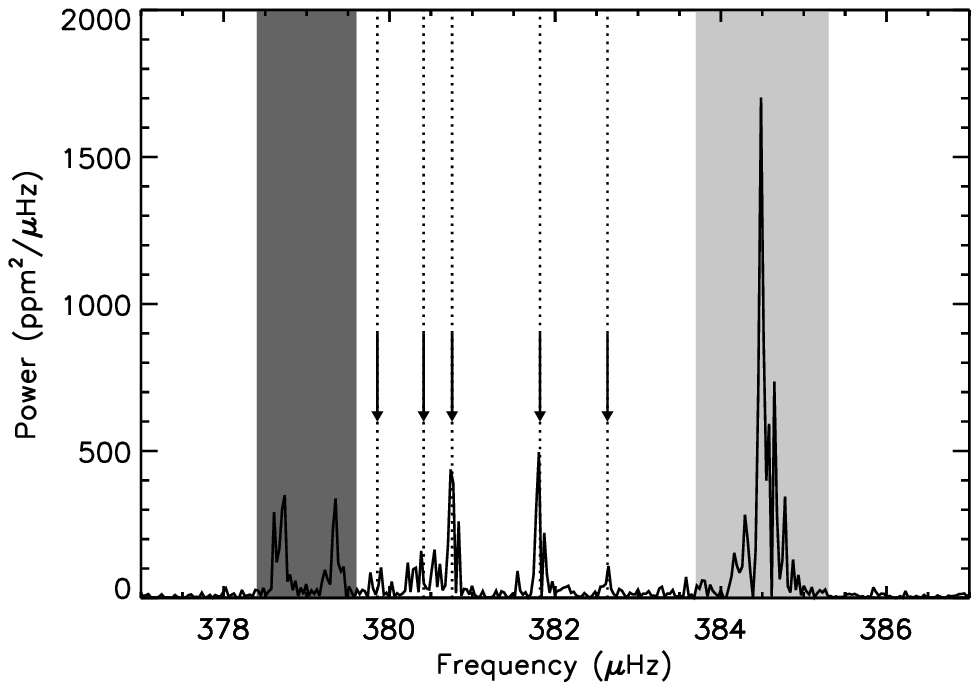}
\end{center}
\caption{Section of the power spectrum of \cible\ in the neighborhood of the 384-$\mu$Hz radial mode (light grey area). The dark grey area corresponds to a rotationally-split $l=1$ mode and the vertical arrows indicate peaks that are attributed to $l=2$ modes (see text).
\label{fig_avcross_l2}}
\end{figure}

The third of these overtones (in the neighborhood of the radial mode at a frequency of 384.5 $\mu$Hz) shows some peculiar features. The $l=2$ mode in this overtone is expected to be located between the radial mode and an $l=1$ mixed mode at a frequency of 379 $\mu$Hz. Fig. \ref{fig_avcross_l2} shows that at least five peaks can be detected in this region instead of three, as would be expected. Besides, when fitting these peaks as the components of one single $l=2$ mode, we find a rotational splitting of about 0.6 $\mu$Hz, which is much too large compared to the splittings of the two other $l=2$ modes (0.13 and 0.18 $\mu$Hz). The most reasonable explanation is that we are in fact detecting an $l=2$ avoided crossing and that two $l=2$ modes are present (one around 380.5 $\mu$Hz and another one around 382.5 $\mu$Hz) instead of one. This hypothesis is plausible since the use of asymptotic relations had predicted an $l=2$ avoided crossing around this frequency (see Sect. \ref{sect_id}) and it will be confirmed by the modeling of the star in Sect. \ref{sect_model}. But even in this case, the multiplets do not have the expected pattern. The frequencies of the components in the multiplets appear to be asymmetrical, contrary to what would be expected for such a slow rotation. There seems to be an asymmetry also in the amplitudes of the components. For these reasons, we failed to obtain rotational splittings for these two $l=2$ modes in avoided crossing. A theoretical study of the interaction between rotation and avoided crossings for $l=2$ modes is under way and could possibly explain the observed profile of the multiplets. However, at this point, we preferred to discard these two modes. \\

%The agreement on the mode parameters between the fitting teams is very poor. For local fits, the obtained inclination angle varies from one team to the other, and even from one mode to the other for each single team. This last point is not necessarily alarming. In main sequence stars, one of the reasons why a global fit was preferred to local fits is that the latter ones tend to converge toward different inclination angles when the signal-to-noise ratio becomes too low. This might be the case for $l=2$ modes in \cible. However, a much bigger concern is that we do not have any model-independent way of identifying $l=2$ avoided crossings in the oscillation spectrum, contrary to $l=1$ modes. As a result, the risk is high to mistake two $l=2$ modes in avoided crossing for the $m=\pm2$ components of an $l=2$ purely p multiplet. This can have serious consequences on the determination of the inclination angle in a global fit. Moreover, this could induce spurious points in our list of rotational splittings. We conclude from these remarks that even tough $l=2$ modes could bring valuable information about the rotation profile in the interior of \cible, the risk of having them contaminate our list of rotational splittings is too high to take them into account at this point. Further comments on the special case of higher degree modes in \cible\ in Sect. \ref{sect_higher_l}.

\begin{figure}
\begin{center}
\includegraphics[width=8cm]{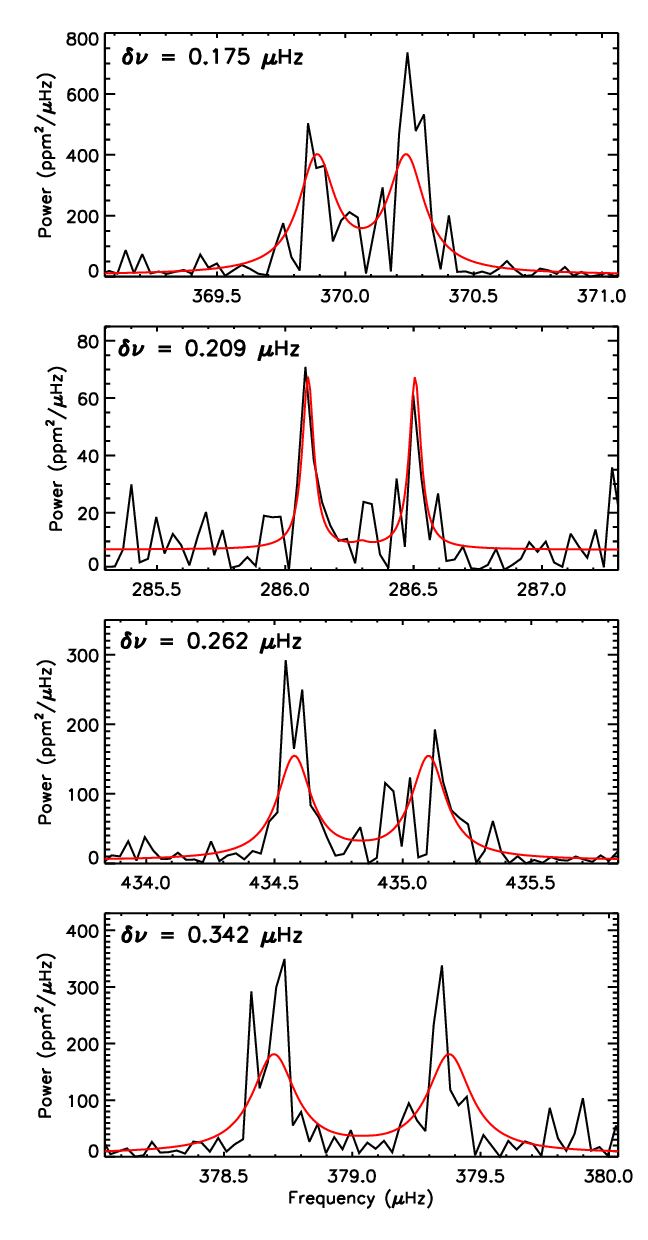}
\end{center}
\caption{Profiles of four $l=1$ multiplets in the oscillation spectrum of \cible. The profiles we obtained when assuming a varying splitting in our fit are overplotted in red. The fitted values of the rotational splittings are specified for each multiplet.
\label{fig_splitvar}}
\end{figure}

\begin{figure}
\begin{center}
\includegraphics[width=8cm]{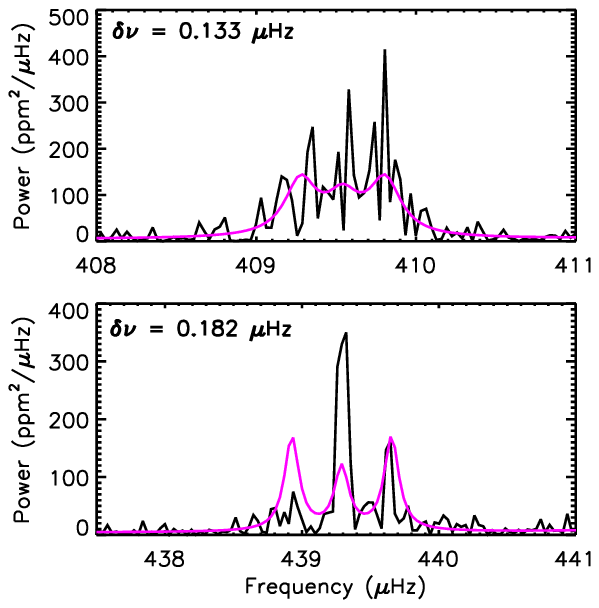}
\end{center}
\caption{Profiles of two $l=2$ multiplets in the oscillation spectrum of \cible. The profiles we obtained when assuming a varying splitting in our fit are overplotted in purple. The fitted values of the rotational splittings are specified for each multiplet.
\label{fig_fitl2}}
\end{figure}

A last global fit was performed, using only the modes for which at least six of the seven fitting teams agreed to estimate the inclination angle of the star. The idea was to remove the possible influence of spurious non-radial modes when determining the angle. We thus obtained an inclination angle of $i=85\pm5^\circ$ and a robust, reliable list of rotational splittings (given in Table \ref{tab_splittings}) that can be safely used to derive information about the star's rotation profile. The obtained splittings range from 0.13 to $0.41\,\mu$Hz, with error bars of $0.03\,\mu$Hz on average. It is therefore clear that there are significant variations of the rotational splitting from one mode to another. To illustrate this, Fig. \ref{fig_splitvar} shows the profiles of four $l=1$ multiplets that have different rotational splittings. The profiles of the two $l=2$ multiplets for which the rotational splitting could be estimated are plotted in Fig. \ref{fig_fitl2}. This shows that the interior of the star rotates differentially in radius. To investigate this further, it is necessary to find a stellar model in order to establish the relation between the obtained rotational splittings and the trapping of the modes.

\begin{table}
\begin{center}
\caption{Estimates of the frequencies and rotational splittings of the detected modes for \cible, obtained by fitting Lorentzian functions to the mode profiles. The rotational splittings are given only for the modes for which at least six of the seven teams agreed to within 1-$\sigma$.
\label{tab_splittings}}
\begin{tabular}{c c c}
\hline \hline
\T \B $l$ & $\nu_{n,l}$ ($\mu$Hz) & $\delta\nu_{n,l}$ ($\mu$Hz) \\
\hline
\T 0 & $271.150\pm0.072$ & n.a.$^\sharp$ \\
0 & $299.057\pm 0.101$ & n.a. \\
0 & $327.239\pm0.051$ & n.a. \\
0 & $355.869\pm0.040$ & n.a. \\
0 & $384.498\pm0.030$ & n.a. \\
0 & $413.478\pm0.035$ & n.a. \\
0 & $442.596\pm0.040$ & n.a. \\
0 & $472.016\pm0.065$ & n.a. \\
0 & $501.328\pm0.068$ & n.a. \\
\B 0 & $531.517\pm0.202 $ & n.a. \\
\hline
\T 1 & $286.297\pm0.015$ & $0.209\pm0.015$ \\
1 & $315.401\pm0.022$ & $0.230\pm0.021$ \\
1 & $333.950\pm0.051$ & $0.327\pm0.041$ \\
1 & $341.592\pm0.017$ & $0.232\pm0.018$ \\
1 & $349.600\pm0.067$ & $0.413\pm0.063$ \\
1 & $361.255\pm0.076$ & $0.363\pm0.087$ \\
1 & $370.063\pm0.023$ & $0.175\pm0.024$ \\
1 & $379.036\pm0.024$ & $0.342\pm0.023$ \\
1 & $392.160\pm0.019$ & $0.252\pm0.019$ \\
1 & $400.559\pm0.022$ & $0.156\pm0.024$ \\
1 & $412.584\pm0.012$ & $0.323\pm0.013$ \\
1 & $425.766\pm0.038$ & $0.216\pm0.040$ \\
1 & $434.839\pm0.023$ & $0.262\pm0.022$ \\
1 & $450.966\pm0.034$ & $0.292\pm0.032$ \\
1 & $460.377\pm0.036$ & $0.270\pm0.036$ \\
1 & $476.696\pm0.072$ & $0.285\pm0.063$ \\
1 & $488.244\pm0.044$ & $0.243\pm0.045$ \\
%1 & $503.998\pm0.060$ & $0.260\pm0.051^\sharp$ \\
1 & $517.382\pm0.058$ & - \\ %$0.400\pm0.052^\sharp$ \\
1 & $534.012\pm0.096$ & - \\ %$0.457\pm0.079^\sharp$ \\
\B 1 & $547.940\pm0.187$ & - \\
\hline
\T 2 & $295.212\pm0.300$ & - \\
2 & $323.649\pm0.218$ & - \\
2 & $352.237\pm0.079$ & - \\
2 & $409.541\pm0.033$ & $0.133\pm0.018$ \\
2 & $439.290\pm0.016$ & $0.182\pm0.016$ \\
2 & $468.457\pm0.087$ & - \\
2 & $498.239\pm0.089$ & - \\
\B 2 & $528.407\pm0.374$ & - \\
\hline
\T 3 & $420.000\pm0.109$ & - \\
\B 3 & $449.455\pm0.114$ & - \\
\hline
\hline 
\end{tabular}
\\
{\small $^\sharp$n.a.: not applicable.}
\end{center}
%{\small \textbf{Notes.} $^{(\star)}$ The metallicity is  defined as $[Z/X]\equiv\log\left[(Z/X)/(Z/X)_{\odot}\right]$.}
\end{table}

%The case of $l=2$ is far more complicated. As mentioned in Sect. \ref{sect_id}, the identification method that we applied to the $l=1$ modes cannot be used for higher degree modes. As a result, we cannot have a prior knowledge of avoided crossings could occur in the power spectrum. One ensuing risk is to mistake two $l=2$ modes in avoided crossing for the $m=\pm2$ components of an $l=2$ purely p multiplet. This could have serious consequences one the determination of the inclination angle in a global fit. Moreover, this would induce spurious rotational splittings in our list of fitted modes. We conclude from these remarks that even tough $l=2$ modes could bring valuable information about the rotation profile in the interior of \cible, the risk of having them contaminate our list of rotational splittings is too high to take them into account at this point. For this reason, we focus exclusively on the $l=1$ modes in the following. We however give further comments on the special case of higher degree modes in \cible\ in Sect. \ref{sect_higher_l}. 

\section{Seismic modeling of \cible \label{sect_model}}

To obtain information about $\Omega(r)$,  we need to have access to the rotational kernels of the modes and therefore to a model of the star. %A thorough modeling of the star is clearly out of the scope of this paper. 
Our aim here is to find a model offering a reasonable agreement with both the atmospheric constraints of the star and the observed mode frequencies, so that we can use the rotational kernels of this model to estimate the rotation profile of the star. 

\subsection{Modeling red giant stars using the frequencies of mixed modes \label{sect_method_model}}

It is known that the modeling of stars with avoided crossings is complex because the timescale of these phenomena is very short compared to the evolution timescale (\citealt{lanzarote}). To remedy this, \cite{deheuvels11} proposed a method specifically designed to handle these stars. They showed that the combined knowledge of the mean large separation $\langle\Delta\nu\rangle$ of the star and the frequency of an avoided crossing $\nu\ind{cross}$ provides extremely precise estimates of the stellar mass and age, for a given physics. The reason for this is that both $\langle\Delta\nu\rangle$ and $\nu\ind{cross}$ are monotonic functions of age. As a result, there is one and only one value of the mass and age for which the observed values of these quantities are simultaneously reproduced. For each considered physics, \cite{deheuvels11} therefore suggested to fix the stellar mass and age to the values required by $\langle\Delta\nu\rangle$ and $\nu\ind{cross}$, which solves the problem caused by the short timescale of avoided crossings.

\begin{figure}
\begin{center}
\includegraphics[width=8cm,clip=1]{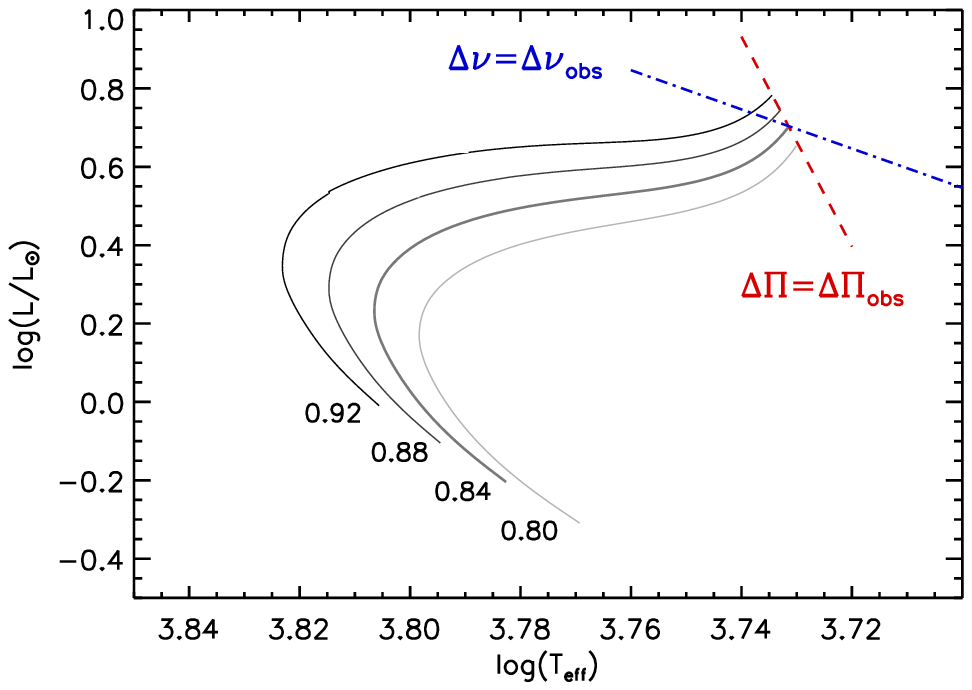}
\end{center}
\caption{Location in the HR diagram of the models that reproduce the observed value of $\Delta\Pi_1$ (red dashed line) and the observed value of $\Delta\nu$ (blue dotted line). The models represented here are computed with $[$Fe/H$]=-1$ dex. The evolutionary tracks of several models, whose masses are specified in $M_\odot$, were overplotted.
\label{fig_deltanu_deltap}}
\end{figure}

This method was tailored for subgiants, whose spectra usually contain few avoided crossings. The star \cible\ is obviously more evolved than these objects because we had to introduce about 15 $\gamma$ modes of degree $l=1$ to account for the observed oscillation spectrum of the star (see Sect. \ref{sect_id}). Consecutive avoided crossings are close to each other and it is hard to isolate them. The method proposed by \cite{deheuvels11} therefore had to be adapted. 

Contrary to subgiants, the $\gamma$ modes of \cible\ have a high enough order to follow the asymptotic theory. Their periods are thus approximately equally spaced by the period spacing, defined as
\begin{linenomath*}
\begin{equation}
\Delta\Pi_l = \frac{\pi}{L} \left( \int_{r_1}^{r_2} \frac{N\ind{BV}}{r} \hbox{d}r \right)^{-1}.
\label{eq_deltap_asymp}
\end{equation}
\end{linenomath*}
where $L\equiv\sqrt{l(l+1)}$ and $N\ind{BV}$ the \vaisala\ frequency. Since $N\ind{BV}$ keeps increasing with age due to the growing central density, the period spacing $\Delta\Pi_l$ monotonically decreases as the star evolves. Therefore, in the method proposed by \cite{deheuvels11}, the quantity $\nu\ind{cross}$ can legitimately be replaced by $\Delta\Pi_1$. The observed values of $\langle\Delta\nu\rangle$ and $\langle\Delta\Pi_1\rangle$ can be used to obtain good first estimates of the stellar mass and age, for any given physics. This point is illustrated in Fig. \ref{fig_deltanu_deltap}. The mass and age can then be fine-tuned to reproduce at best the observables of the star. We note that since $\langle\Delta\nu\rangle$ and $\langle\Delta\Pi_1\rangle$ are used only to derive a first estimate of the stellar parameters, we can safely use the values derived for these quantities in Sect. \ref{sect_id}.

%As is illustrated in Fig. \ref{fig_deltanu_deltap}, the observed values of $\langle\Delta\nu\rangle$ and $\langle\Delta\Pi_1\rangle$ can be used to obtain first estimates of the stellar mass and age, for any given physics. For this purpose, we have been using the values derived of $\langle\Delta\nu\rangle$ and $\langle\Delta\Pi_1\rangle$ that were derived in Sect. \ref{sect_id}. For each considered physics, the stellar mass and age was then fine-tuned to match the observed mode frequencies 

\subsection{Modeling of \cible}

We applied the method described above to model the star \cible.

\subsubsection{Properties of the models}

All the models were computed with the evolution code \cesam\ (\citealt{cesam}). We used the OPAL 2005 equation of state and opacity tables as described in \cite{lebreton08}. The nuclear reaction rates were computed using the NACRE compilation (\citealt{angulo99}). The atmosphere was described by Eddington's grey law. We assumed the classical solar mixture of heavy elements of \cite{grevesse93}. Convection was treated using the Canuto-Goldman-Mazzitelly (CGM) formalism (\citealt{canuto96}). This description involves a free parameter, the mixing length, which is taken as a fraction $\alpha\ind{CGM}$ of the pressure scale height $H_p$. In this work, we assumed a value of $\alpha\ind{CGM}$ calibrated on the Sun ($\alpha_\odot=0.64$, \citealt{samadi06}). The effects of microscopic diffusion were neglected in this study.
%, but the effect of a change in this parameter is addressed in Sect. \ref{sect_}. The effect of microscopic diffusion are neglected in this study. 

We used the oscillation code LOSC (\citealt{losc}) to compute the mode frequencies of the models. It is well known that absolute mode frequencies are affected by our improper modeling of surface convection (see e.g. \citealt{christensen97}). 
We used the correction of surface effects proposed by \cite{kjeldsen08}, which consists of adding to the mode frequencies a power law whose exponent is calibrated on the Sun. With our evolution code and treatment of convection, we found an exponent of 4.25, which was used for all the models in this work. We also note that some modes in the spectrum of the star are mixed and are therefore less sensitive to surface effects. To take this into account, the surface correction of non-radial modes was multiplied by a factor $Q_{n,l}^{-1}$, where $Q_{n,l}$ corresponds to the ratio of the mode inertia to the inertia of the closest radial mode, as prescribed by \cite{asteroseismology} (Chap. 7). We note that the empirical way of correcting for surface effects proposed by \cite{kjeldsen08} is probably not optimal, and efforts are currently made to take them into account in a more satisfactory way (e.g. \citealt{gruberbauer12}).

\subsubsection{Results \label{sect_model_results}}

\begin{table*}
  \begin{center}
  \caption{Parameters of the best-fit models as a function of the assumed metallicity. \label{tab_param_model}}
\begin{tabular}{c c c c c c}
\hline \hline
\T \B Model & A & B & C & D & E \\
\hline
\T \B [Z/X] (dex) &      -0.75 &       -1. &       -1.25 &       -1.5 &       -1.75 \\
\hline
\T Mass ($M_\odot$) &       0.880 &       0.836 &       0.804 &       0.790 &       0.770 \\
Age (Gyr) & 11.3 & 12.2 & 13.1 & 13.4 & 14.3 \\
$T\ind{eff}$ (K) &        5245 &        5363 &        5452 &        5521 &        5566 \\
Radius ($R_\odot$) &        2.67 &        2.62 &        2.59 &        2.58 &        2.55 \\
$\log g$ &        3.527 &        3.520 &        3.514 &        3.510 &        3.508 \\
\B $r\ind{BCE}^\sharp$ ($R_\star$) &       0.33 &       0.36 &       0.39 &       0.43 &       0.46 \\
\hline 
\T $\chi^2\ind{atm}$ & 0.3 & 0.2 & 0.5 & 1.0 & 1.3 \\
\B $\chi^2\ind{seis}$ & 7.7 & 5.5 & 5.6 & 6.8 & 9.6 \\
\hline
\end{tabular} \\
$^\sharp$ BCE: base of the convective envelope
\end{center}
\end{table*}
%[Z/X] (dex)                            -0.75     -1     -1.25     -1.5     -1.75
%chi2_atm(Teff_photo)              1.2       0.5       0.4        0.6       0.8
%chi2_atm(Teff_spectro)            0.1       1.0       2.8        4.7       6.1
%chi2_sismo                              7.7       5.5       5.6        6.8       9.6

%\begin{figure}
%\begin{center}
%\includegraphics[width=9cm,clip=1]{fig_modele_colloq.ps}
%\end{center}
%\caption{\'Echelle diagram of the mode frequencies of \cible\ obtained in Sect. \ref{sect_mle} (filled black circles). We overplotted the mode frequencies of the best-fit model obtained with $[Z/X]=-0.05$ dex (blue diamonds: $l=0$ modes, red triangles: $l=1$ modes, green squares: $l=2$ modes). The size of the symbols has been chosen to decrease logarithmically as the mode inertia increases.
%\label{fig_echelle_model}}
%\end{figure}

\begin{figure}
\begin{center}
\includegraphics[width=8cm]{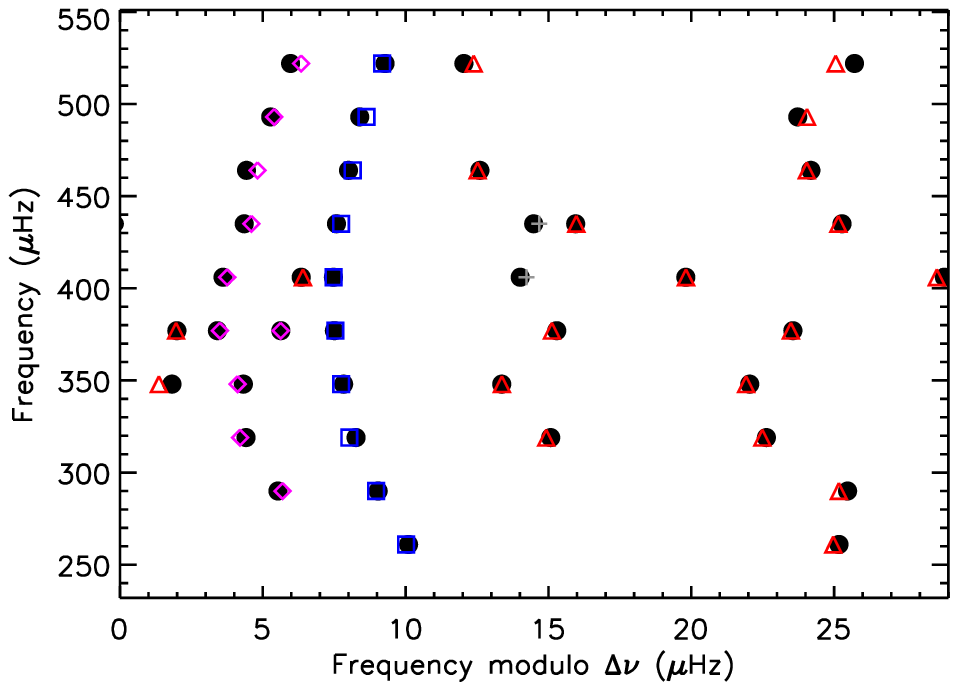}
\end{center}
\caption{\'Echelle diagram of the mode frequencies of \cible\ obtained in Sect. \ref{sect_mle} (filled black circles). We overplotted the mode frequencies of the best-fit model obtained with $[Z/X]=-1$ dex (blue squares: $l=0$ modes, red triangles: $l=1$ modes, purple diamonds: $l=2$ modes, grey plusses: $\ell=3$ modes).
\label{fig_echelle_model}}
\end{figure}

We applied the method described in Sect. \ref{sect_method_model} to find stellar models reproducing both the seismic and atmospheric constraints of \cible. Since the metallicity of the star is rather uncertain, we tried several values of $(Z/X)$ ranging from $-1.75$ to $-0.75$ dex. The initial helium content was estimated from galactic evolution, by supposing a ratio $\Delta Y/\Delta Z=3\pm2$ (\citealt{pagel98}) and an primordial helium abundance of $Y\ind{p}=0.2477$ (\citealt{peimbert07}). For each value, we determined the stellar mass and age for which we both have $\langle\Delta\nu\rangle=28.9\pm0.2\,\mu$Hz and $\langle\Delta\Pi_1\rangle=107.1\pm2.3$ s, as was obtained for the star in Sect. \ref{sect_id}. We then fine-tuned these parameters to match the atmospheric constraints ($T\ind{eff}$ and $\log g$) and the frequencies of the observed modes as closely as possible. We have here used our photometric estimate of the effective temperature ($T\ind{eff}=5470\pm150$ K, see Sect. \ref{sect_photo}), but we verified that our conclusions remain unchanged if we consider our spectroscopic estimate of $T\ind{eff}$ instead. The agreement between the models and the observations was estimated using the reduced $\chi^2$ function defined as
\begin{linenomath*}
\begin{equation}
\chi^2 = \frac{1}{N} \sum_{k=1}^N \frac{\left(\mathcal{O}\ex{obs}_k-\mathcal{O}\ex{mod}_k\right)^2}{\sigma_k^2}
\end{equation}
\end{linenomath*}
where $\mathcal{O}\ex{obs}_k$, $k=1,N$ correspond to the $N$ observables available for the star, $\sigma_k$ their error bars, and $\mathcal{O}\ex{mod}_k$ the corresponding values in the computed models. The number of seismic constraints is much larger than the number of atmospheric constraints. To avoid drowning the contribution of the latter in the total value of the $\chi^2$, we computed a separate $\chi^2$ for the seismic observables ($\chi^2\ind{seis}$) and for the atmospheric observables ($\chi^2\ind{atm}$).

The properties of the best-fit models that we obtained for each considered value of the stellar metallicity are given in Table \ref{tab_param_model}. The observables are best reproduced for a metallicity around $-1$ dex. The low value of $\chi^2\ind{atm}$ and $\chi^2\ind{seis}$ for these models indicates a close agreement with both the atmospheric and the seismic constraints of the star, which is confirmed by Fig. \ref{fig_echelle_model}. For lower abundances of heavy elements, the effective temperature of the models differs from the observations (this is all the more true when considering our spectroscopic estimate of $T\ind{eff}$ for the star) and more importantly, the agreement with the observed mode frequencies deteriorates. \cible\ is found to be a low-mass star (0.77 to 0.88 $M_\odot$) at the bottom of the red giant branch. Its age (ranging from 11.3 to 14.3 Gyr) is consistent with that of an old halo star. The surface gravity of the models matches the seismic estimate of $\log g=3.55\pm0.03$ that was obtained in Sect. \ref{sect_sismo} within 1-$\sigma$ error bars. We note that in all our best-fit models, there is an $l=2$ avoided crossing around 380 $\mu$Hz (see Fig. \ref{fig_echelle_model}). This confirms the hypothesis made in Sect. \ref{sect_mle}, that two $l=2$ mixed modes are detected in the oscillation spectrum of the star around this frequency. Finally, the spectroscopic upper limit that we obtained on $v\sin i$ ($<1\pm1$ \kms, see Sect. \ref{sect_spectro}) combined with the estimate of the star's radius that we get from the models (about 2.6 $R_\odot$) enables us to set an upper limit on the surface rotation rate of $\Omega\ind{surf}<88\pm90$ nHz (we recall that the inclination angle of the star has been found to be $85\pm5^\circ$).

\begin{figure*}
\begin{center}
\includegraphics[width=8cm]{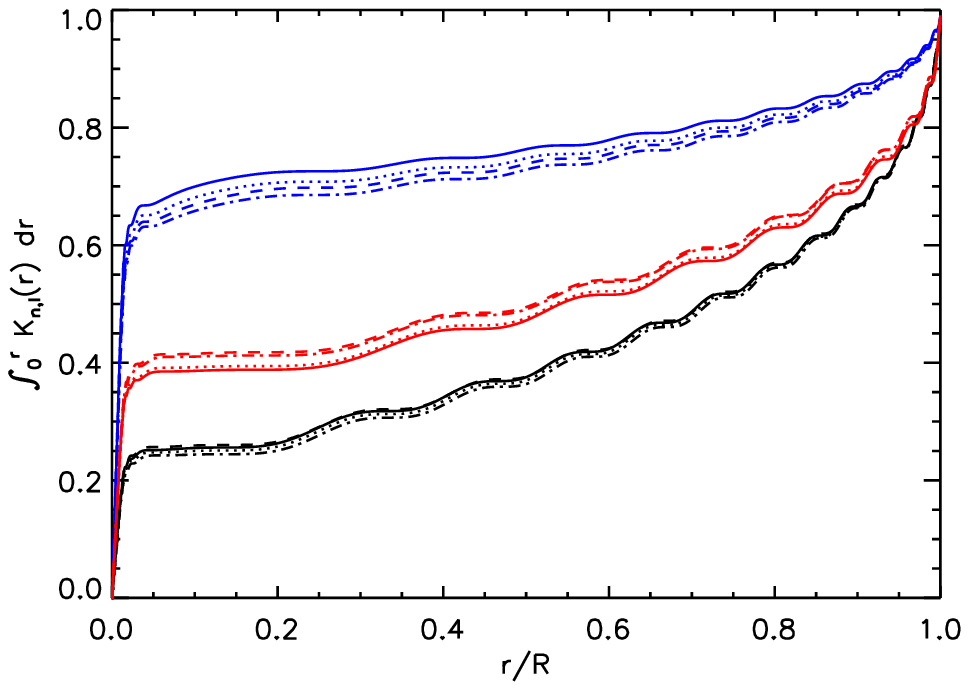}
\includegraphics[width=8cm]{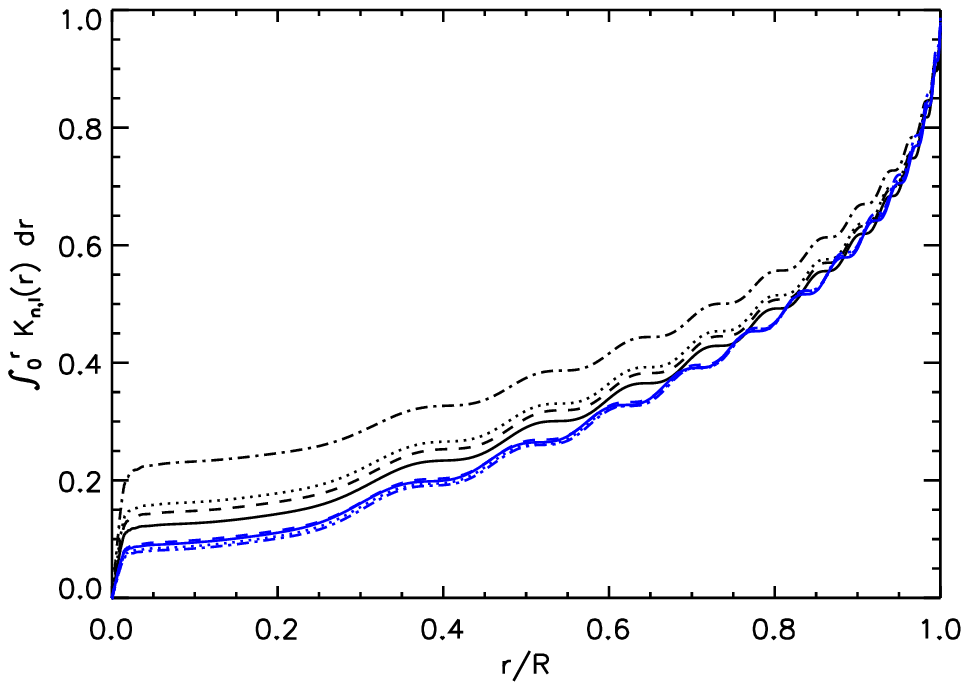}
\end{center}
\caption{\textbf{Left: }Cumulative normalized integral of the rotational kernels of three $l=1$ modes that were detected in the spectrum of \cible, computed using models A (full lines), B (dotted lines), C (dashed lines), and D (dash-dot lines). The black curves correspond to a mode mainly trapped in the envelope, the blue curves to a mode mainly trapped in the core and the red curves to a mode in between. \textbf{Right:} Same as left panel for the two $l=2$ modes for which the rotational splitting could be determined (black: $\nu=409\,\mu$Hz, blue: $\nu=439\,\mu$Hz).
%The full line corresponds to a mode trapped mainly in the envelope ($q=0.35$) and the dashed line to a mode trapped in the core ($q=5.05$).
\label{fig_kernel}}
\end{figure*}

From the models, we can compute the rotational kernels of the detected modes. Fig. \ref{fig_kernel}a shows the integrated kernels of three $l=1$ modes trapped in different regions inside the star, computed using the models presented in Table \ref{tab_param_model} (only model E was excluded because its seismic $\chi^2$ is larger). The variations in the kernel profiles are quite small when switching from one model to another, which shows that the trapping of the $l=1$ modes depends only weakly on the model we choose for the star. The integrated kernels of the two $l=2$ modes for which the rotational splitting could be measured are plotted in Fig. \ref{fig_kernel}b. For one of them ($\nu=439\,\mu$Hz), we reach the same conclusion as for the $l=1$ modes. However, for the second mode ($\nu=409\,\mu$Hz), the variations are larger. The reason for this is that this mode is in fact undergoing an avoided crossing with another $l=2$ mode trapped mainly in the core. Unfortunately, this latter mode could not be detected. As a result, the trapping of the detected mode is uncertain and varies from one model to the other, which generates the differences that we observe between the rotational kernels of the models. We note that a longer data set will perhaps make it possible to detect the gravity-dominated $l=2$ mode, which would solve this problem. The $l=1$ modes are not prone to this kind of uncertainty because they can be detected even when they are trapped essentially in the core.

\subsection{Rotational splitting vs mode trapping}

Having access to a stellar model of the star, we were then able to relate the observed rotational splittings to the trapping of the modes, the idea being to search for the signature of differential rotation in the interior of the star. For this purpose, we introduced for each mode the weighted radius $r_{n,l}$
\begin{linenomath*}
\begin{equation}
r_{n,l} \equiv \frac{\int_0^R r K_{n,l}(r) \, \hbox{d}r}{\int_0^R K_{n,l}(r) \, \hbox{d}r}
\label{eq_rloc}
\end{equation}
\end{linenomath*}
which corresponds to the mean location of the rotational kernel inside the star. Small values of $r_{n,l}$ indicate that the mode is trapped in the core and therefore more g-like, whereas if $r_{n,l}$ is close to 1, the mode is trapped in the envelope and close to an acoustic mode.

%To describe the trapping of the modes, we used the ratio $q$ between the kinetic energy of the mode in the g-mode cavity and the kinetic energy in the p-mode cavity, i.e.
%\begin{linenomath*}
%\begin{equation}
%q \equiv \frac{\int_{r\ind{a}}^{r\ind{b}} \rho r^2 (\xi_r^2+L^2\xi_h^2)\,\hbox{d}r}{\int_{r\ind{c}}^{r\ind{d}} \rho r^2 (\xi_r^2+L^2\xi_h^2)\,\hbox{d}r}
%\label{eq_EgEp}
%\end{equation}
%\end{linenomath*}
%where $r\ind{a}$ and $r\ind{b}$ are the turning points of the g-mode cavity, $r\ind{c}$ and $r\ind{d}$ those of the p-mode cavity. A large value of $q$ indicates that the mode is trapped in the core whereas if $q$ is small, the mode is trapped in the envelope.

%\begin{figure}
%\begin{center}
%\includegraphics[width=9cm]{fig_kernel_l2.ps}
%\end{center}
%\caption{Integrated rotational kernels of the two $l=2$ modes for which the rotational splitting could be determined (black: $\nu=409\,\mu$Hz, blue: $\nu=439\,\mu$Hz). The kernels are computed using either model A (full lines) or model B (dashed lines).
%%The full line corresponds to a mode trapped mainly in the envelope ($q=0.35$) and the dashed line to a mode trapped in the core ($q=5.05$).
%\label{fig_kernel}}
%\end{figure}

\begin{figure}
\begin{center}
\includegraphics[width=8cm]{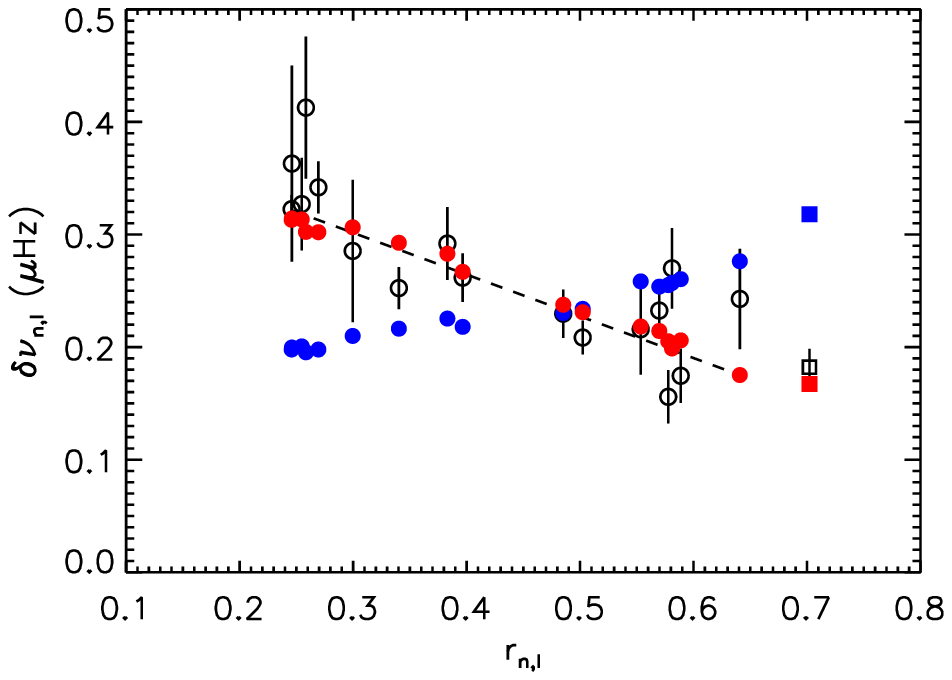}
\end{center}
\caption{Rotational splittings of the modes as a function of the weighted radii $r_{n,l}$ given by Eq. \ref{eq_rloc}. The empty symbols ($l=1$: circles, $l=2$: squares) correspond to the observations with 1-$\sigma$ error bars, the filled blue symbols to an optimal solid-body rotation model and the filled red symbols to an optimal two-zone model.
\label{fig_rloc_split}}
\end{figure}

In Fig. \ref{fig_rloc_split}, we plotted the measured splittings as a function of the weighted radii $r_{n,l}$, that were computed for each detected mode using model B as a reference model. We deliberately omitted the $l=2$ mode around $\nu=409\,\mu$Hz in this figure because it is the only mode for which the value of the ratio $q$ depends significantly on the choice of the reference model, for the reason we mentioned in Sect. \ref{sect_model_results}. We observe a very clear correlation between these quantities: the modes which have a g-mode behavior tend to have a larger splitting than the ones which behave as p modes. This clearly suggests that the core of \cible\ rotates faster than the envelope.

\section{Inversion of the rotational profile $\Omega(r)$ \label{sect_inversion}}

We established that the core of \cible\ is rotating faster than the surface. In order to quantify this differential rotation, we then tried to estimate the rotation profile of the star by inverting Eq. \ref{eq_split_inv}, using  the observed splittings and the rotational kernels of the modes computed in Sect. \ref{sect_model}. We start by recalling general background on inversions (see \citealt{christensen90} for more details).

We performed linear inversions, which means that for each radius $r_0$ inside the star, the inferred rotation profile $\bar{\Omega}(r_0)$ can be expressed as a combination of the rotational splittings
\begin{linenomath*}
\begin{equation}
\bar{\Omega}(r_0) = \sum_{k=1}^M c_k(r_0)\delta\omega_k
\label{eq_comblin}
\end{equation}
\end{linenomath*}
where we have used a subscript $k=1,M$ for the detected modes instead of their radial order $n$ and degree $l$, for convenience. The coefficients $c_k(r_0)$ are either determined during the process of the inversion or they can be calculated separately. The solution is characterized by the \textit{averaging kernels} defined as
\begin{linenomath*}
\begin{equation}
\mathcal{K}(r;r_0)\equiv\sum_{k=1}^M c_k(r_0) K_k(r)
\label{eq_kernel_av}
\end{equation}
\end{linenomath*}
where $K_k(r)$ is the rotational kernel of the $k\ex{th}$ mode. By combining Eq. \ref{eq_split_inv} and \ref{eq_comblin}, we obtain that $\bar{\Omega}(r_0)=\int_0^R \mathcal{K}(r;r_0)\Omega(r) \,\hbox{d}r$. We thus aim at getting the averaging kernel $\mathcal{K}(r;r_0)$ as localized as possible around $r_0$ to ensure that $\bar{\Omega}(r_0)$ is a good approximation of the true rotation rate $\Omega(r_0)$. In all the cases below, the averaging kernels are required to have unit integral, so that the solution $\bar{\Omega}(r_0)$ constitutes a proper average of the true rotation rate and its standard deviation is then given by
\begin{linenomath*}
\begin{equation}
\sigma_{\bar{\Omega}(r_0)} = \sqrt{\sum_{k=1}^M \left[ c_k(r_0) \sigma_{\delta\omega,k} \right]^2}
\label{eq_err_omega}
\end{equation}
\end{linenomath*}
where the $\sigma_{\delta\omega,k}$ are the standard deviations of the observed modes.

We have tried two of the most commonly used inversion techniques to invert the rotation profile of the star, the Regularized Least Squares (RLS) method and the Optimally Localized Averages (OLA) method. These methods were both successfully applied to estimate the internal rotation profile of the Sun (e.g. \citealt{schou98}, \citealt{chaplin99}). In what follows, we have used the mode rotational kernels computed from model B, which is the one that offers the closest fit to the observations, and we have excluded from our list of measured splittings the $l=2$ mode around $409\,\mu$Hz because its rotational kernel is too model-dependent. The results that are described are very similar if we use models A, C or D instead of model B. 

%Two different approaches are possible. We can combine the rotational kernels of the modes computed in Sect. \ref{sect_model} with simple rotation profiles $\Omega(r)$ to obtain theoretical rotational splittings through the relation
%\begin{linenomath*}
%\begin{equation}
%\delta\omega_{n,l} = \int_0^R K_{n,l}(r) \Omega(r) \, \hbox{d}r
%\label{eq_inverse}
%\end{equation}
%\end{linenomath*}
%These theoretical splittings can then be confronted to the observed splittings (\textit{forward method}). The alternative is to invert Eq. \ref{eq_inverse} in order to estimate the rotation profile of \cible\ from the observed rotational splittings (\textit{inverse method}). Both methods were attempted in the present work.

%In the following, we have used the mode rotational kernels computed from model B, which is the one that offers the closest fit to the observations, and we have excluded from our list of measured splittings the $l=2$ mode around $409\,\mu$Hz because its rotational kernel is too model-dependent. The results that are described are very similar if we use models A, C or D instead of model B. 

\subsection{Least Squares methods}

One approach to inverting the rotation profile of the star is to try to reproduce as closely as possible the observed rotational splittings by performing a Least-Squares fit to the observations. The objective is thus to minimize the $\chi^2$ function
\begin{linenomath*}
\begin{equation}
\chi^2 = \sum_{k=1}^M \frac{\left[\delta\omega_k-\int_0^R \bar{\Omega}(r)K_k(r) \,\hbox{d}r\right]^2}{\sigma_{\delta\omega,k}^2}
\label{eq_chi2}
\end{equation}
\end{linenomath*}
In practice, this can be done by discretizing the rotation profile $\bar{\Omega}(r)$ on an $N$-point grid. The stellar radius is split into $N$ regions delimited by the radii $0=r_0<\hdots<r_N=1$ and for $j=1,N$ we set
\begin{linenomath*}
\begin{equation}
\bar{\Omega}(r) = \Omega_j, \;\;\; \hbox{for} \;\;\; r_{j-1}<r\leqslant r_j
\end{equation}
\end{linenomath*}
In the following, we first tried small values of $N$, resulting in simple profiles. Then, for larger values of $N$, the solution needs to be regularized to ensure numerical stability.

%We tried several simple rotation profiles $\Omega(r)$ for \cible\ and each time adjusted their parameters to reproduce at best the observed rotational splittings of the modes. 

\subsubsection{Solid-body rotation profiles ($N=1$)}

The case $N=1$ corresponds to a uniform rotation rate $\bar{\Omega}(r)=\Omega_1$ throughout the star. By minimizing the $\chi^2$ function given by Eq. \ref{eq_chi2}, we obtained an optimal value of $\Omega_1=328\pm7$ nHz. The agreement with the observed data is very poor (blue circles in Fig. \ref{fig_rloc_split}), yielding a reduced $\chi^2$ of about 17. This clearly shows that we can reject the hypothesis of a solid-body rotation inside the star. Fig. \ref{fig_rloc_split} shows that if we assume a uniform rotation, the modes trapped in the core are expected to have slightly smaller splittings than the modes trapped in the envelope --- this is caused by the fact that the kernels of p modes have approximately unit integral, while the integral of g-mode kernels is smaller than one. In the observations, it is the contrary, which confirms that the core rotates faster than the envelope.

\subsubsection{Two-zone models ($N=2$) \label{sect_2zone}}

For $N=2$, the star is split into two uniformly rotating regions separated at a radius $r_1$, with $\bar{\Omega}(r)=\Omega_1$ toward the core and $\bar{\Omega}(r)=\Omega_2$ toward the surface.

\begin{figure}
\begin{center}
\includegraphics[width=8cm]{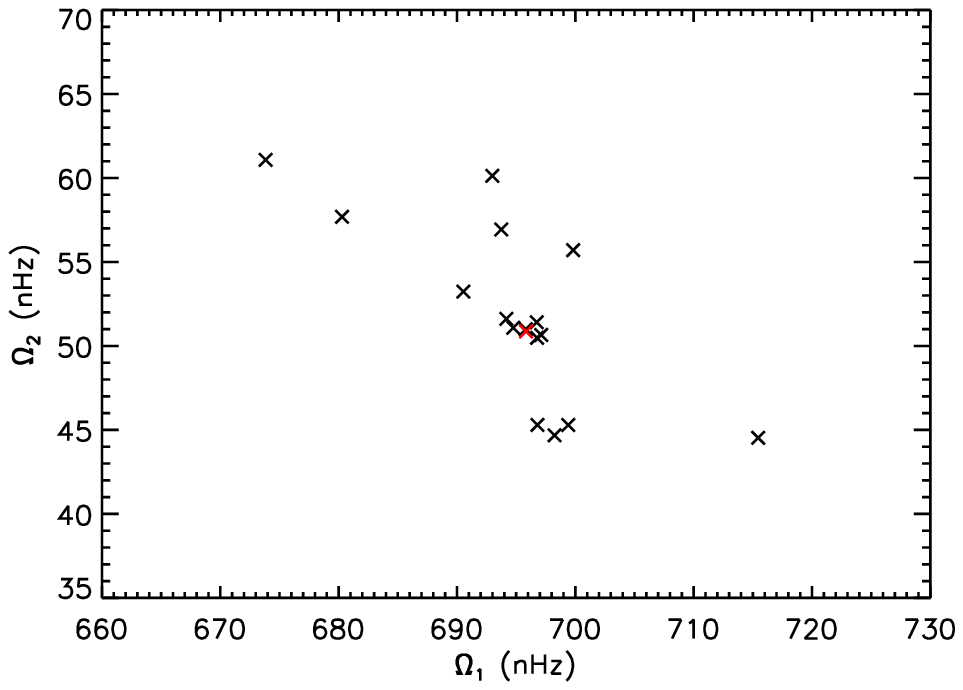}
\caption{Values of $\Omega_1$ and $\Omega_2$ obtained when rejecting points one by one in the rotational splittings list (black crosses). The red cross indicates the result obtained when considering all the rotational splittings.}
\label{fig_reject}
\end{center}
\end{figure}

\begin{figure}
\begin{center}
\includegraphics[width=8cm]{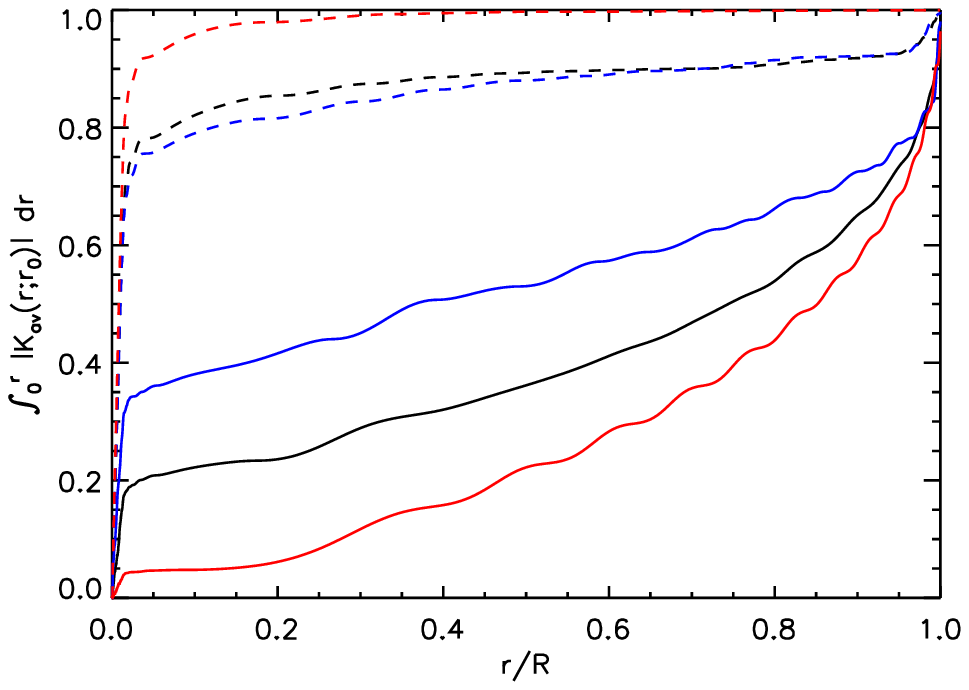}
\caption{Integrated modulus of the averaging kernels of the core (dashed lines) and the surface (full lines) obtained by describing the rotation profile as a two-zone model (black), by using the RLS inversion method with a 100-point grid (blue) or the OLA inversion method with the same grid (red). For more clarity, the integrals were normalized to their surface value.}
\label{fig_kernel_av}
\end{center}
\end{figure}

We first separated these zones at the interface $r\ind{CE}$ between the convective envelope and the radiative core. We note that based on the model we consider, this value of $r\ind{CE}$ varies (see Table \ref{tab_param_model}). However, the results that we obtained for the rotation rates vary little. By minimizing the $\chi^2$ function, we obtained $\Omega_1=696\pm24$ nHz and $\Omega_2=51\pm19$ nHz. The agreement between the theoretical splittings and the observed ones is much better than with the solid-body profile, as can be seen in Fig. \ref{fig_rloc_split}. The reduced $\chi^2$ is now about 1.6, which indicates a fairly good agreement with the observations. To estimate how robust this result is, we repeated the same procedure 18 times, rejecting points one by one in our list of measured rotational splittings. Fig. \ref{fig_reject} shows that the values of $\Omega_1$ and $\Omega_2$ remain quite stable in all cases. We then calculated the averaging kernels corresponding to the solution that we obtained. Instead of plotting the averaging kernels themselves, we chose to represent in Fig. \ref{fig_kernel_av} the cumulative integral of their modulus in order to estimate more efficiently to which region of the star they are most sensitive (we recall that unlike the rotational kernels of the modes, the averaging kernels are not necessarily positive functions). Apart from a small contribution from the surface, the core kernel is quite well localized. It is especially sensitive to the innermost 2\% of the star in radius. On the contrary, the envelope kernel appears to be severely contaminated by the core.

%\begin{linenomath*}
%\begin{equation}
%\left( \begin{array}{c c}
%\sum_k \frac{I_{k,1}^2}{\sigma_{\delta\omega,k}^2} & \sum_k \frac{I_{k,1}I_{k,2}}{\sigma_{\delta\omega,k}^2} \\
%\noalign{\medskip}
%\sum_k \frac{I_{k,1}I_{k,2}}{\sigma_k^2} & \sum_k \frac{I_{k,2}^2}{\sigma_{\delta\omega,k}^2}
%\end{array} \right)
%\left( \begin{array}{c}
%\Omega_1 \vphantom{\frac{I_{k,2}^2}{\sigma_{\delta\omega,k}^2}} \\
%\noalign{\medskip}
%\Omega_2 \vphantom{\frac{I_{k,2}^2}{\sigma_{\delta\omega,k}^2}} 
%\end{array} \right)
%= \left( \begin{array}{c}
%\sum_k \frac{I_{k,1}\delta\omega_k}{\sigma_{\delta\omega,k}^2} \\
%\noalign{\medskip}
%\sum_k \frac{I_{k,2}\delta\omega_k}{\sigma_{\delta\omega,k}^2}
%\end{array} \right)
%\label{eq_2zone}
%\end{equation}
%\end{linenomath*}

%\textbf{By solving Eq. \ref{eq_2zone}, we can find a set of coefficients $c_{k,j}$ such that the optimal rotation rate of the $j\ex{th}$ zone ($j=1,2$) can be written as $\Omega_j=\sum_k c_{k,j}\delta\omega_k$. From these coefficients, we can build averaging kernels for the core and the envelope
%\begin{equation}
%\mathcal{K}_j(r) = \sum_{k=1}^M c_{k,j}K_k(r)
%\end{equation}
%Since $\Omega_j=\int_0^R \mathcal{K}_j(r) \Omega(r) \, \hbox{d}r$, we would like the averaging kernel $\mathcal{K}_j(r)$ to be as localized as possible inside the $j\ex{th}$ region, so that $\Omega_j$ is a good approximation of the average rotation rate in this zone. Contrary to the rotational kernels themselves, the averaging kernels are not necessarily positive functions. We plotted in Fig. \ref{fig_kernel_av} the integrated modulus of both averaging kernels. Apart from a small contribution from the surface, the core kernel is quite well localized. On the contrary, the envelope kernel appears to be significantly contaminated by the core.}

\begin{figure*}
\begin{center}
\includegraphics[width=8cm]{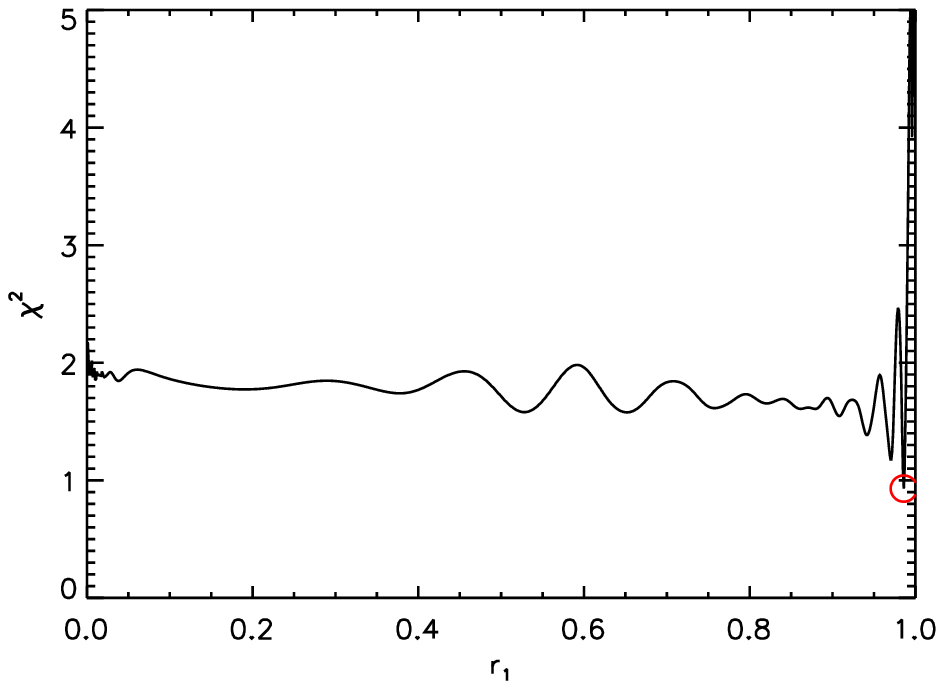}
\includegraphics[width=8cm]{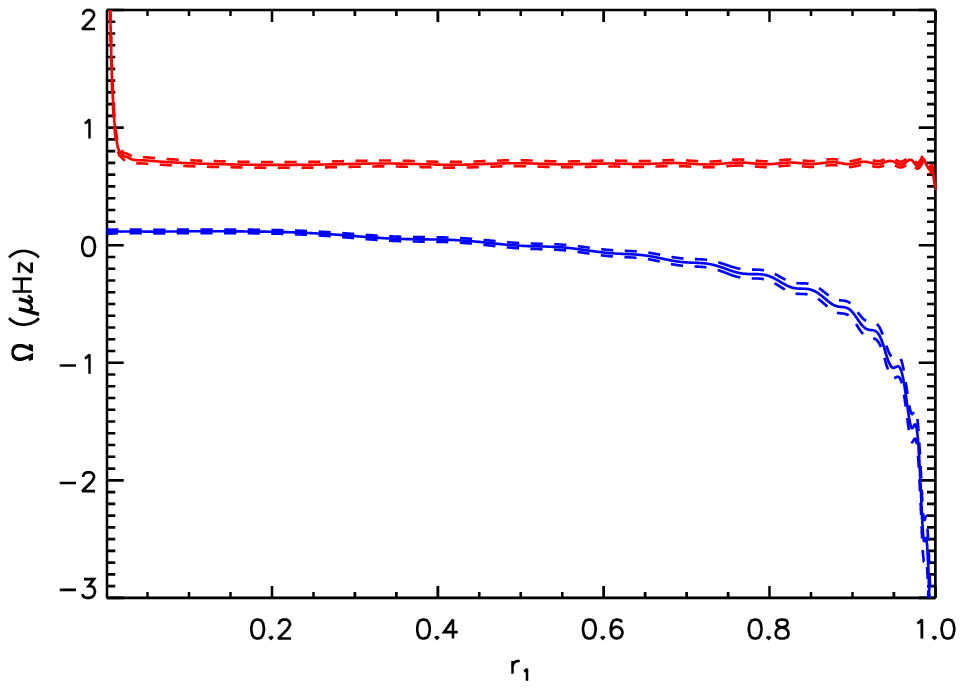}
\caption{\textbf{Left}: Variations in the reduced $\chi^2$ value as a function of the radius $r_1$. The red circle indicates the minimum of the function. \textbf{Right}: Optimal values of the core rotation $\Omega_1$ (red) and the surface rotation $\Omega_2$ (blue) as a function of $r_1$. The dashed lines indicate the error bars.}
\label{fig_chi2_rc}
\end{center}
\end{figure*}

We also tried to consider the intermediate radius $r_1$ as a free parameter instead of fixing it to $r\ind{CE}$. In this case, the minimization is not as trivial since $\nabla\chi^2$ is not a linear function of the rotation rates at the grid points. However, the simplicity of the profile allowed us to vary $r_1$ from 0 to 1 and each time determine the optimum values of $\Omega_1$ and $\Omega_2$. Fig. \ref{fig_chi2_rc} shows the variations of the reduced $\chi^2$ that we obtained as a function of $r_1$. The first remark is that the minimum $\chi^2$ is as low as $0.93$, significantly smaller than the one obtained with $r_1=r\ind{CE}$. The corresponding rotation profile is very strange. We obtained $r_1=0.985$, $\Omega_1=730\pm27$ nHz and $\Omega_2=-2513\pm178$ nHz, i.e. a flat rotation in the largest part of the star and a thin layer at the surface spinning fast in the opposite direction. 

\begin{figure}
\begin{center}
\includegraphics[width=8cm]{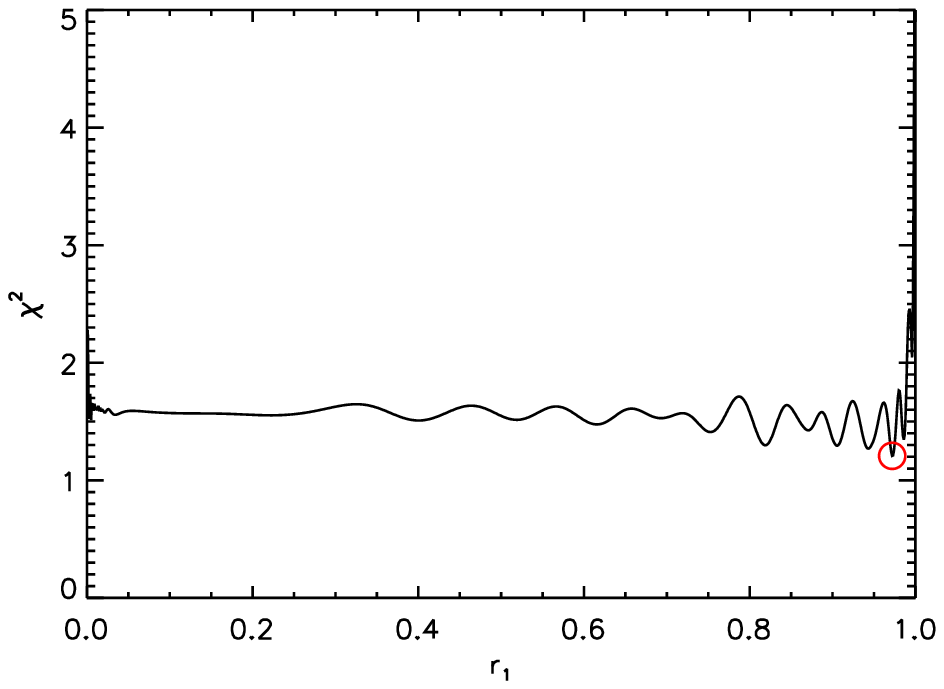}
\caption{Same as the left panel of Fig. \ref{fig_chi2_rc}, but for a simulation (see text).}
\label{fig_chi2_rc_simu}
\end{center}
\end{figure}

However, there are several indications showing that this solution should be discarded. First, the $\chi^2$ function has an oscillatory behavior and thus many secondary minima throughout the star's interior. Secondly, knowing that the radius of the star is about 2.6 $R_\odot$, a surface rotation rate of $\Omega_2=-2513$ nHz would require a surface velocity of about 29 \kms, clearly inconsistent with the spectroscopic upper limit of $v\sin i<1\pm1$ \kms. Finally, we performed a simulation to show that the minimum at $r_1=0.985$ is spurious. We computed theoretical rotational splittings using the optimal rotation profile that had been obtained when setting $r_1=r\ind{CE}$ (=0.36 for model B). We added to these theoretical values a Gaussian noise whose width corresponds to the error bars of the observed splittings. We then tried to recover the input value of $r_1$ by applying the same procedure as before. We performed several iterations of this procedure, which all led to a $\chi^2$ function very similar to the one that we obtained from the data. They indeed have an oscillatory behavior and their minimum can be found anywhere between $r=0$ and $r=1$ instead of recovering the input value $r\ind{CE}=0.36$. Fig. \ref{fig_chi2_rc_simu} shows the example of an iteration for which the minimum of the $\chi^2$ function is close to the surface. All this shows that we can ignore this strange counter-rotating profile.

%\subsubsection{Continuous profiles \label{sect_linear}}

%We also tried several continuous profiles for $\Omega(r)$. The simplest of those is a linear function of the type
%\begin{linenomath*}
%\begin{equation}
%\Omega(r) = \Omega_1 + \alpha r
%\end{equation}
%\end{linenomath*}
%Finding the optimal values of $\Omega_1$ and $\alpha$ is straightforward, following the same method as in Sect. \ref{sect_2zone}. We obtained $\Omega_1=705\pm25$ nHz and $\alpha=-761\pm35$ nHz, meaning that the surface would be weakly counter-rotating ($\Omega\ind{s}=-56\pm25$ nHz). The reduced $\chi^2$ we obtained is 1.6, which is similar to the one of the two-zone model with $r_0=r\ind{CE}$.

%Other more sophisticated continuous profiles were tried, but they did not improve the agreement with the observations, in spite of the increase in the number of free parameters to describe $\Omega(r)$.

%\subsection{An attempt to invert $\Omega(r)$}

%Until now, the only star for which the quality of the seismic data was high enough to invert the rotation profile from the rotational splittings is the Sun (\citealt{schou98}). For \cible, we have precise determination of the splittings for 15 modes whose rotational kernels are sensitive to different parts of the star because they are not trapped in the same regions. We thus took this opportunity to study the feasibility of an inversion of the rotation profile in the interior of the star. We tried two of the most currently used techniques to perform our inversions, the Regularized  Least Squares method (RLS) and the Optimally Localized Averages method (OLA).

\subsubsection{Regularized Least Squares method}

We then explored the case of larger values of $N$. In this case, the finite number of measured splittings for the star is obviously not sufficient to reconstruct the whole function $\Omega(r)$, which makes any solution non unique. Besides, the solutions are very sensitive to both the discretization and the measurement errors $\sigma_{\delta\omega,k}$. As a result, the problem is ill-conditioned and needs to be regularized. The Regularized Least Squares method consists in adding to the classical $\chi^2$ function to be minimized a well chosen regularization function.

%Our aim is to find an estimate of the rotation profile $\Omega(r)$ based on the measurements of the rotational splittings that we obtained in Sect. \ref{sect_mle}. If the problem were well-posed, a traditional least-squares fit could be applied and we would seek to minimize the $\chi^2$ function defined as
%\begin{linenomath*}
%\begin{equation}
%\chi^2\equiv \sum_{k=1}^M \frac{\left[\delta\omega_k-\int_0^R K_k(r)\Omega(r)\, \hbox{d}r\right]^2}{\sigma_{\delta\omega,k}^2}.
%\label{eq_chi2_RLS}
%\end{equation}
%\end{linenomath*}
%However, the finite number of measured splittings for the star is obviously not sufficient to reconstruct the whole function $\Omega(r)$, which makes any solution non unique. Besides, the solutions are very sensitive to both the discretization and the measurement errors $\sigma_{\delta\omega,k}$. As a result, the problem is ill-posed and needs to be regularized. The Regularized Least Squares method consists in adding to the classical $\chi^2$ function to be minimized a well chosen regularization function. This method was successfully applied to estimate the internal rotation profile of the Sun (e.g. \citealt{schou98},\citealt{chaplin99}).

To apply the RLS method to our star, we followed the procedure prescribed by \cite{christensen90}. The regularization term was chosen as a smoothness constraint  on the solution. We thus minimized the function
\begin{linenomath*}
\begin{equation}
J\equiv\chi^2+\mu u F(\Omega)
\label{eq_J}
\end{equation}
\end{linenomath*}
where $F(\Omega)$ is the regularization function, which was taken as the norm of the second derivative of $\Omega(r)$, i.e. $F(\Omega)\equiv\|\Omega''(r)\|^2$. The factor $\mu$ is a trade-off parameter between the smoothness of the solution and the minimization of the $\chi^2$ function. For more convenience, we added a normalization factor $u$ defined as $u\equiv\left(\sum_k \sigma_{\delta\omega,k}^2/M\right)^{-1}$. The minimization of the function $J$ can be written as a linear problem (\citealt{christensen90}), which can readily be solved to estimate the optimal rotation profile.

To determine a satisfactory value of the trade-off parameter $\mu$, we generated artificial rotation profiles $\Omega(r)$, from which we computed theoretical rotational splittings. We then applied the method described above with different values of the regularization parameter $\mu$ to try to recover the input rotation profile. We found that a value of $\mu\approx100$ provides a good compromise. Smaller values of $\mu$ introduce spurious large-amplitude oscillations in the recovered profile, while larger values of $\mu$ lead to an over-regularization and the obtained profile corresponds to the one that minimizes $F(\Omega)$ (a straight line if the norm of the second derivative of $\Omega(r)$ is taken as a smoothness condition). We therefore used $\mu=100$ when inverting $\Omega(r)$ from the observed splittings.

\begin{figure*}
\begin{center}
\includegraphics[width=8cm]{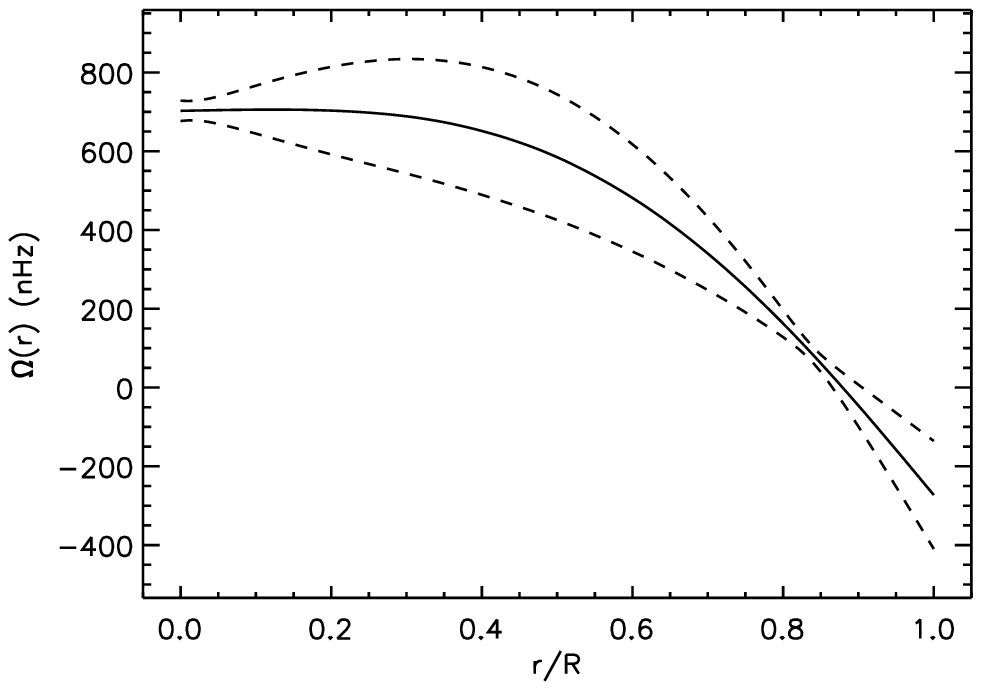}
\includegraphics[width=8cm]{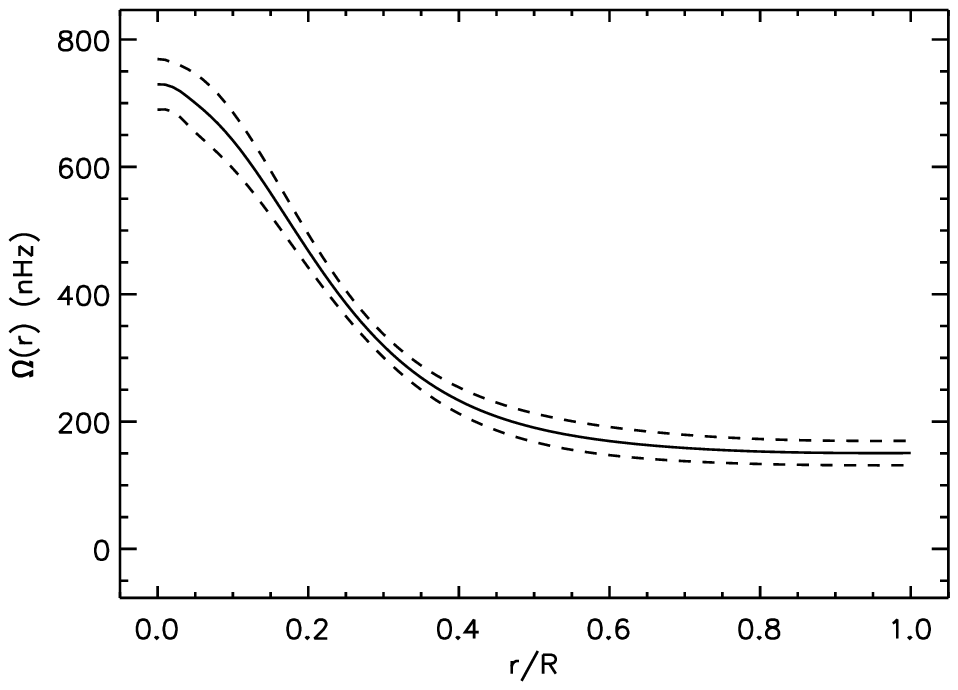}
\end{center}
\caption{\textbf{Left:} Inverted rotation profile obtained with the RLS method (100 points, regularization with the second derivative and $\mu=100$). The dashed lines indicate 1-$\sigma$ error-bars. We note that apart from in the core, the averaging kernels are ill-localized and the inverted rotation rates are not reliable (see text). \textbf{Right:} Same as left panel using the OLA method.
\label{fig_omega_RLS}}
\end{figure*}

Fig. \ref{fig_omega_RLS} shows the rotation profile that we obtained by discretizing $\Omega(r)$ over a 100-point grid. We found a rotation rate in the core of $695\pm29$ nHz, very similar to the one derived with two-zone rotation models. A rotation rate of $-272\pm137$ nHz was obtained at the surface of the star. The agreement between the rotational splittings obtained with the inverted rotation profile and the observed ones is only slightly better than the one we reached with a two-zone model in Sect. \ref{sect_2zone}. Useful information about the quality of the inverted profile can be obtained by computing the corresponding averaging kernels. Details on how these kernels are calculated with the RLS method are given by \cite{christensen90}. An inspection of these averaging kernels shows that only the one located in the core has a satisfactory shape. It is in fact very similar to the core kernel obtained with the two-zone model. It is well localized in the innermost 1.4\% of the stellar radius (17\% in mass) and has a small contribution from the surface (see Fig. \ref{fig_kernel_r0}). On the contrary, all the other kernels have large leakage from both the core and the surface. The integrated modulus of the surface kernel, for instance, has a large contribution from the core (see Fig. \ref{fig_kernel_av}). Only the estimate of the rotation rate obtained for the core is thus reliable. In particular, the slight counter-rotation that we obtained at the surface cannot be regarded as significant.

\subsection{Optimally Localized Averages method}

Instead of seeking to reproduce at best the observed rotational splittings like the RLS method, the Optimally Localized Averages (OLA) method consists in calculating localized averages of the true rotation profile in different regions of the star. For this purpose, the method builds for each point $r_j$ inside the star a linear combination of the mode kernels $\sum_k c_k(r_j)K_k(r)$ such that the resulting averaging kernel $\mathcal{K}(r;r_j)$ is as close as possible to a Dirac function centered on $r_j$.

Following \cite{backus68}, we searched for the coefficients $c_k(r_j)$ by minimizing the function
\begin{linenomath*}
\begin{equation}
J=12 \int_0^R \mathcal{K}(r;r_j)^2 (r-r_j)^2\,\hbox{d}r + \mu \sum_{k=1}^M \left[ c_k(r_j)\sigma_{\delta\omega,k} \right]^2
\end{equation}
\end{linenomath*}
with the constraint that $\int \mathcal{K}(r;r_j)\,\hbox{d}r=1$, for each point $r_j$; $\mu$ is a trade-off parameter between resolution of the averaging kernels and error magnification.

%\begin{figure}
%\begin{center}
%\includegraphics[width=9cm]{fig_omega_OLA_l2.ps}
%\caption{Inverted rotation profile obtained with the OLA method with a ten-point grid. The dashed lines indicate 1-$\sigma$ error-bars.}
%\label{fig_omega_OLA}
%\end{center}
%\end{figure}

\begin{figure*}
\begin{center}
\includegraphics[width=8cm]{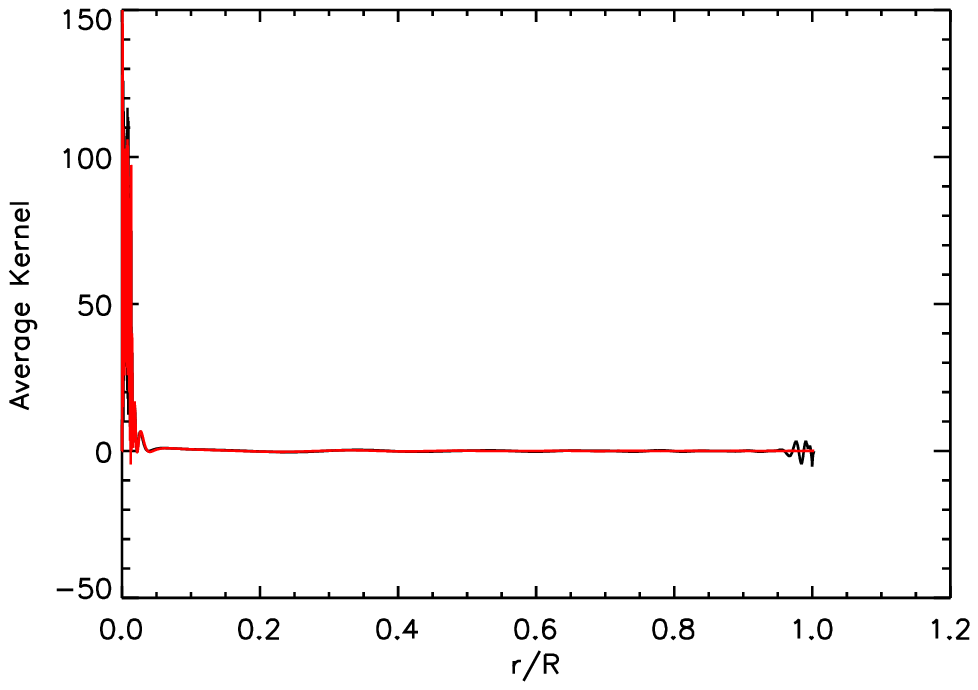}
\includegraphics[width=8cm]{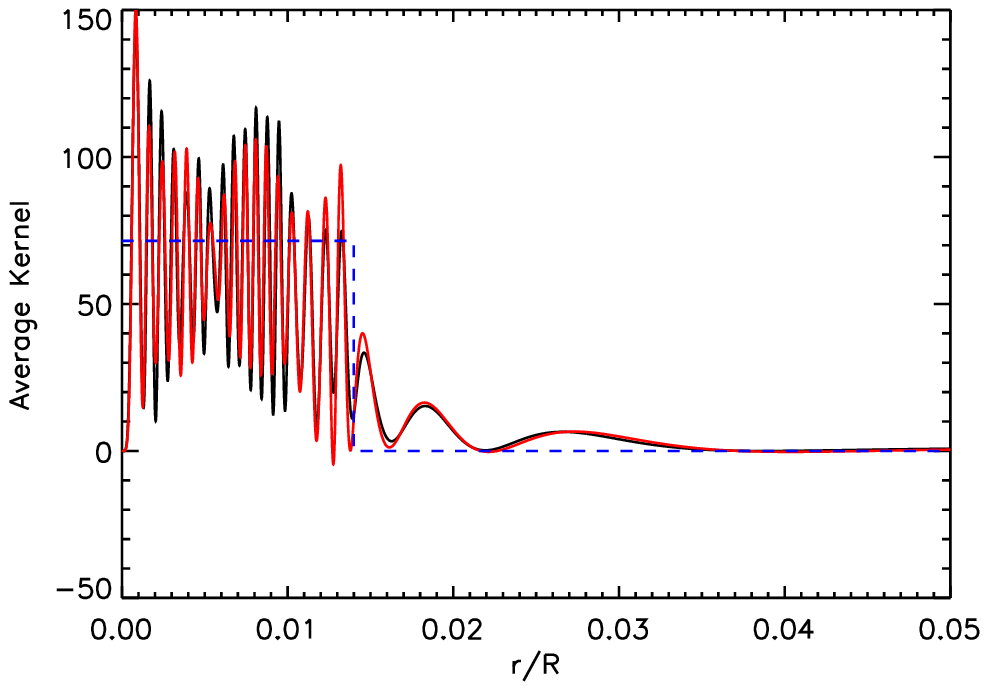}
\caption{\textbf{Left}: Averaging kernel in the center of the star obtained with the RLS method (black curve) and the OLA method (red curve). \textbf{Right}: Zoom of the left panel in the core. The blue dashed line corresponds to the step function $H(r)$ (see text).}
\label{fig_kernel_r0}
\end{center}
\end{figure*}

In our case, it is very hard to obtain localized averaging kernels, so we chose to set $\mu=0$. Even this way, most averaging kernels suffer from very large leakage from other parts of the star (especially from the core and the surface). The core kernel is an exception (see Fig. \ref{fig_kernel_r0}). Fig. \ref{fig_kernel_av} shows that the leakage from the surface in the core kernel that was observed with the RLS method has now been cancelled. If we assume that the rotation profile of the star varies smoothly in the core, then the core kernel can be well approximated by a step function $H(r)$ between 0 and a small radius $\tilde{r}\approx0.014\,R$ (see Fig. \ref{fig_kernel_r0}). This means that
\begin{linenomath*}
\begin{equation}
\bar{\Omega}(0) \approx \int_0^R \frac{H(r)}{R} \Omega(r) \hbox{d}r = \frac{\int_0^{\tilde{r}} \Omega(r) \hbox{d}r}{r_0}
\end{equation}
\end{linenomath*}
The quantity $\bar{\Omega}(0)$ is therefore a very good approximation of the average of $\Omega(r)$ in the innermost 1.4\% in radius, which corresponds to 17\% in mass, of the star. We obtain $710\pm51$ nHz for this average, which is consistent with the result of the RLS method.

Fig. \ref{fig_kernel_av} shows the integrated value of the surface averaging kernel $|\mathcal{K}(r;1)|$ along the stellar radius. The contribution from the core to this kernel has been significantly decreased compared to the RLS method. However, there remains a small leakage that can have significant consequences because we know that the core rotation is larger than the envelope rotation. The estimate of the surface rotation ($150\pm19$ nHz) is therefore certainly overestimated and only gives an upper limit of the surface rotation.

\section{Conclusion}

In this paper, we obtained a precise seismic determination of the rotation rate in the core of the early red giant \cible\ and we proved that it spins at least five times faster than the surface.

\cible\ is a low-mass evolved star, which is located at the bottom of the red giant branch. Solar-like oscillations have been detected in the oscillation spectrum of the star, derived from one year of \textit{Kepler} observations (quarters Q5-6-7-8). Due to the evolution stage of the star, many of its non-radial modes have a mixed nature, which means that they behave both as g modes in the core and as p modes in the envelope. We found that many of these mixed modes are very clearly split by stellar rotation and we therefore set out to probe the rotation profile of the star. 

We performed a seismic analysis of the oscillation spectrum of the star and were able to determine precisely the rotational splittings of 19 $l=1$ and $l=2$ modes. They were found to range from 0.13 to 0.41 $\mu$Hz, with error bars of $0.03\,\mu$Hz on average, thereby suggesting that the interior of the star is differentially rotating in radius. We then found a stellar model reproducing very well both the atmospheric and the seismic properties of the star. We used this model to study the relation between the observed rotational splittings and the regions in the star where the modes are trapped. We found a clear correlation between these quantities that unambiguously indicated that the core rotates faster than the envelope in \cible. 

Finally, we performed inversions of the rotation profile of the star, using the observed splittings and the rotational kernels of our best-fit models. We used both the RLS (Regularized Least Square) and the OLA (Optimally Localized Averages) methods and obtained the following results:
\begin{itemize}
\item We were able to determine a very robust and precise estimate of the core rotation of the star. \textit{All the methods} that we used (RLS, OLA) provided a core rotation rate consistent with $\Omega\ind{c}=710\pm51$ nHz within 1-$\sigma$ error bars. Besides, we obtained similar values when using the rotational kernels of other models of the star computed with different metallicities, so this result seems to be only weakly model-dependent. We showed, using the OLA method, that this rotation rate in fact corresponds to a very good approximation of the average of $\Omega(r)$ in the innermost 1.4\% of the stellar radius. It is ironic that the core rotation rate of \cible\ could be measured while the solar core rotation is still uncertain for $r < 0.2R_\odot$ (\citealt{chaplin99}).
\item We obtained an upper limit for the surface rotation of $\Omega\ind{s}<150\pm19$ nHz. This enabled us to establish that the core rotates at least five times faster than the surface in this star.
\end{itemize}

We note that in this study we essentially focused on the rotational splittings of $l=1$ modes. This is partly caused by the SNR of modes of degree $l\geqslant2$, which in most cases remains too low to reliably determine their rotational splittings. This problem should be at least partially solved by the growing data set from the \textit{Kepler} spacecraft (this star will continue to be on the short-cadence target list at least through Q12 and hopefully for the remainder of the mission). However, we also showed that the profiles of certain $l=2$ mixed modes split by rotation significantly differ from the expected one and it is our opinion that some theoretical work still remains to be done to better understand the effects of rotation on mixed modes of degree $l\geqslant2$.

We are entering a new era in the study of the transport of angular momentum in stars because we now have access to observational constraints on the internal rotation profiles of stars, which were longed for since a very long time. We will very likely find among the \textit{Kepler} targets other subgiants and red giants whose internal rotation can be inferred by interpreting the rotational splittings of mixed modes. For instance, if we assume that the rotation is nearly rigid at the end of the main sequence (as it is in the Sun), then the differential rotation observed in subgiants and red giants is entirely caused by the core contraction in the post main sequence stage. The ratio between the core rotation and the surface rotation at different luminosities along the giant branch should bring very valuable constraints on the timescale of the exchange of angular momentum between the core and the envelope. This should help us determine which mechanisms of angular momentum transport dominate and how efficient they are.

%We insist that the seismic study of subgiants observed with \textit{Kepler} is only starting. The oscillation spectra of many more of them will be analyzed shortly and we can hope to detect the signature of rotation in mixed modes for several of them. The present work definitely shows that if we find such objects, we can expect to derive quantitative estimates of the rotation in their interior.

%This work also opens many interesting perspectives to investigate about the transport of angular momentum in stars. It should be very instructive to confront the results of this study to the rotation profiles that are predicted by the state of the art models of transport of angular momentum for a star such \cible. For instance, we know that when evolved stars are moving toward and climbing up the red giant branch, their core contracts while their envelope expends. The rotation in the 
%interior of the star is likely to be affected by these phenomena. The resulting rotation profile should strongly depend on the coupling between the core and the envelope and stars like \cible\ are very good candidates to constraint this coupling.

\begin{acknowledgements}
The authors are very grateful to the \textit{Kepler} team for building such a marvelous mission and providing exquisite data for seismology. Funding for this Discovery mission is provided by NASA's Science Mission Directorate. We wish to thank the KITP at UCSB for their warm hospitality during the research program "Asteroseismology in the Space Age". This KITP program was supported in part by the National Science Foundation of the United States under Grant No. NSF PHY05–51164. This work was supported in part by NSF grant AST-1105930 (SD, SB). WJC and YE acknowledge financial support from the UK Science and Technology Facilities Council (STFC). LG and TS acknowledge support from the German Science Foundation under SFB 963 “Astrophysical Flow Instabilities and Turbulence”. DRR acknowledges financial support through a postdoctoral fellowship from the ``Subside fédéral pour la recherche 2011'', University of Liège. This research was partially supported by grant AYA2010-17803 from the Spanish National Research Plan (CR). SH acknowledges financial support from the Netherlands Organisation of Scientific Research (NWO). NCAR is partially funded by the National Science Foundation.
\end{acknowledgements}

\bibliographystyle{aa.bst} % style aa.bst
\bibliography{biblio} % your references Yourfile.bib

\begin{appendix}

\section{Modeling of the background in the Power Spectral Density \label{app_bg}}

The background $B(\nu)$ is modeled as the sum of a white noise $B\ind{w}$ describing the photon noise (which dominates the spectrum at high frequency) and a Harvey profile $B\ind{g}(\nu)$ modeling the granulation spectrum. As prescribed by \cite{harvey85}, we assumed for $B\ind{g}(\nu)$ a function of the type
\begin{linenomath*}
\begin{equation}
B\ind{g}(\nu) = \frac{4\sigma\ind{g}^2\tau\ind{g}}{1+(2\pi\nu\tau\ind{g})^{\alpha\ind{g}}}
\end{equation}
\end{linenomath*}
where $\tau\ind{g}$ and $\sigma\ind{g}$ correspond to the characteristic timescale and amplitude of the granulation (see \citealt{mathur11} for a discussion on methodologies to fit the background). The exponent $\alpha\ind{g}$ corresponds to the slope of the power law at high frequency. In most analyses of solar-like oscillation spectra, a function $B(\nu)$ is fitted to the background prior to the fit of the p-mode component. This has the advantage of stabilizing the latter fit and it has been shown several times that this has negligible influence on the results. While fitting the background component, the p-mode contribution is modeled as a Gaussian function. We used an MLE (Maximum Likelihood Estimation) method to fit the background, taking into account the fact that the bins of the PSD follow a 2-degree-of-freedom $\chi^2$ distribution. The fitted background is shown in Fig. \ref{fig_spec} and the obtained parameters are given in Table \ref{tab_bg}. As a by-product, we obtain an estimate of the frequency of the maximum signal $\nu\ind{max}=406\pm3\,\mu$Hz, which corresponds to the centroid of the Gaussian used to describe the stellar pulsation contribution in the power spectrum.

\begin{table}
  \begin{center}
  \caption{Fitted parameters of the background model $B(\nu)$ (see text). \label{tab_bg}}
%\begin{tabular*}{\textwidth}{@{}@{\extracolsep{\fill}} c c c c c c c @{}}
\begin{tabular}{c c c c c c c}
\hline \hline
\T \B $\sigma\ind{g}$ ($10^{-3}$ ppm) & $\tau\ind{g}$ (s) & $\alpha\ind{g}$ & $B\ind{w}$ (ppm$^2/\mu$Hz) \\
%\B (ppm) & (s) & & (ppm$^2/\mu$Hz) & (ppm$^2/\mu$Hz) & ($\mu$Hz) & ($\mu$Hz) \\
\hline
%\T \B $86.4\pm0.6$ &$ 1001\pm21$ & $2.74\pm0.05$ & $2.137\pm0.007$ & $22.5\pm0.5$ & $406\pm1$ & $59\pm1$ \\
%\T \B $86.4\pm0.6$ &$ 1001\pm21$ & $2.74\pm0.05$ & $2.137\pm0.007$ \\
\T \B $99\pm5$ &$ 1096\pm59$ & $2.17\pm0.05$ & $3.32\pm0.04$ \\
\hline 
\end{tabular}
\end{center}
\end{table}

\end{appendix}

\end{document}